\documentclass[prd,apst,tightenlines,preprintnumbers,amsmath,amssymb,showpacs,superscriptaddress,showkeys,notitlepage]
{revtex4-1}

\usepackage{graphicx}
\usepackage{amsmath}
\usepackage{amssymb}
\usepackage{verbatim}   
\usepackage{color}      
\usepackage{subfigure}  
\usepackage{hyperref}   

\newcommand{\be}{\begin{equation}}
\newcommand{\ee}{\end{equation}}
\newcommand{\bea}{\begin{eqnarray}}
\newcommand{\eea}{\end{eqnarray}}
\newcommand{\ba}{\begin{eqnarray*}}
\newcommand{\ea}{\end{eqnarray*}}
\newcommand{\bi}{\begin{itemize}}
\newcommand{\ei}{\end{itemize}}
\newcommand{\nn}{\nonumber}
\newcommand{\Dmq}{\Delta m^2}

\begin{document}

\preprint{ANL-HEP-PR-11-70}
\preprint{EFI 11-32}

\title{Implications of sterile neutrinos for medium/long-baseline neutrino experiments\\ and the determination of $\theta_{13}$.\\}

\author{Bhubanjyoti Bhattacharya}
\author{Arun M. Thalapillil}
\affiliation{Enrico Fermi Institute and Department of Physics, University of Chicago, 5620 South Ellis Avenue, Chicago, IL 60637.}
\author{Carlos E. M. Wagner$^{1,\,}$}
\affiliation{Enrico Fermi Institute and Department of Physics, University of Chicago, 5620 South Ellis Avenue, Chicago, IL 60637.}
\affiliation{KICP and Dept. of Physics, Univ. of Chicago, 5620 S. Ellis
  Ave.,Chicago IL 60637.}
 \affiliation{HEP Division, Argonne National Laboratory, 9700 Cass
  Ave., Argonne, IL 60439.}

\date{\today}
\date{\today}
\begin{abstract}
\par
We revisit some of the recent neutrino observations and anomalies in the context of sterile neutrinos. Among our aims is to understand more clearly some of the analytic implications of the current global neutrino fits from short baseline experiments. Of particular interest to us are the neutrino disappearance measurements from MINOS and the recent indications of a possibly non-vanishing angle, $\theta_{13}$, from T2K, MINOS and Double-CHOOZ. Based on a general parametrization motivated in the presence of sterile neutrinos, the consistency of the MINOS disappearance data with additional sterile neutrinos is discussed. We also explore the implications of sterile neutrinos for the measurement of $|U_{\mu3}|$ in this case. We then turn our attention to the study of $|U_{e3}|$ extraction in electron neutrino disappearance and appearance measurements. In particular, we study the effects of some of the additional CP phases that appear when there are sterile neutrinos. We observe that the existence of sterile neutrinos may induce a significant modification of the $\theta_{13}$ angle in neutrino appearance experiments like T2K and MINOS, over and above the ambiguities and degeneracies that are already present in 3-neutrino parameter extractions. There are reactor experiments, for instance those measuring $\nu_e$ disappearance like Double-CHOOZ, Daya Bay and RENO, where this modification is less significant and therefore the extracted $|U_{e3}|$ value when sterile neutrinos are present is close to the one that would be obtained in the 3-neutrino case. Based on our study, we also conclude that the results from T2K imply a $90 \%$ C.L.  lower-bound on $|U_{e3}|$, in the ``$\,3+2$" neutrino case, which is still within the sensitivity of future reactor neutrino experiments like Daya Bay, and consistent with the one-$\sigma$ range of $\sin^22\theta_{13}$ recently reported by the Double-CHOOZ experiment. Finally, we argue that for the recently determined best-fit parameters, the results in the ``$\,3+1$"  scenario would be very close to the medium/long baseline results obtained in the ``$\,3+2$" case analyzed in this work.
\end{abstract}
\keywords{Neutrino oscillations, sterile neutrinos, long-baseline experiments, reactor angle.}
\pacs{}

\maketitle

\begin{section}{Introduction}
\par
Neutrinos have now been unequivocally established to be massive particles, but with very small masses. Experiments over the past two decades have firmly established a framework of neutrino oscillations that describe solar, atmospheric and reactor neutrino experiments (see for instance \cite{PDG} and references therein).
\par
The experiments are consistent with the existence of three electroweak eigenstates $(\nu_e, \nu_\mu,\nu_\tau)$ and three mass-eigenstates $(\nu_1,\nu_2,\nu_3)$. While the absolute neutrino mass scale has been very difficult to measure, the mass squared differences between the mass-eigenstates ($\Dmq_{21},\,\Dmq_{32}$) are known to good accuracy~\cite{{PDG},{Abe:2010hy},{Adamson:2011ig}}. Also, two of the mixing angles in the lepton sector ($\theta_{12},\,\theta_{23}$) are known to good significance~\cite{{pastexpt1},{Abe:2010hy},{Adamson:2011ig}} and the third ($\theta_{13}$) is being measured and will be measured to better and better accuracy by current and forthcoming experiments~\cite{{Apollonio:2002gd},{Akiri:2011zz},{Wang:2011tp},{Jeon:2011zz}}. It is already clear, for instance, that the mixing in the lepton sector is very distinct from the quark sector.
\par
Inspite of these spectacular successes there are still many outstanding questions related to neutrinos. For instance it is not understood why neutrinos have such tiny masses or why their mixing angles are so much different from the quark sector. There have also been discrepancies from various short-baseline experiments that have been very hard to accommodate in the three active-neutrino picture. 
This has led to many studies incorporating additional singlet neutrino states to the framework~\cite{{Sorel:2003hf},{Nelson:2010hz},{Kopp:2011qd},{Giunti:2011ht},{Barger:2011rc},{Donini:2001xy},{Dighe:2007uf},{Donini:2008wz},{Donini:2007yf},{Meloni:2010zr}}. 
\par
Our main focus in this paper will be to gain a better analytical understanding of scenarios with additional singlet neutrino states and how they may affect current and forthcoming medium/long-baseline neutrino experiments. We are particularly interested in the measurements from MINOS and the determination of the reactor angle $\theta_{13}$ at various medium/long-baseline neutrino experiments. A study similar in spirit to ours was done in \cite{deGouvea:2008qk}, for the case of an additional sterile neutrino. The focus of our study will albeit be different from theirs and will also be motivated by the current global fits, incorporating two additional sterile neutrinos. 
\par
In sections II and III we briefly review the current state of neutrino experiments and the viability of sterile neutrinos in the context of the standard model and cosmology. In section IV we briefly outline the short-baseline limit of neutrino oscillations and the global fits based on them. We also fix our notations here. In section V we study the implications of these short-baseline global fits to medium/long-baseline neutrino experiments and explore various theoretical features. In section VI we give a summary of our results.
\end{section}

\begin{section}{Current state of Neutrino Observations}
Let us briefly review the current state of neutrino parameters. In a three neutrino framework let us label the $\nu$ mass eigenstates by latin indices, $i\,\epsilon\,(1,2,3)$, and electroweak eigenstates by greek indices, $\alpha\,\epsilon\,(e,\mu,\tau)$. 
\par
In the three neutrino framework, various experiments have measured the two mass-squared differences to be~\cite{{PDG},{Abe:2010hy},{Adamson:2011ig}}
\bea
|\Delta m^{2}_{32}|~&\simeq&~2.4\times10^{-3}~~{\text eV}^{2}\; , \\ \nn
|\Delta m^{2}_{21}|~&\simeq&~7.6\times10^{-5}~~{\text eV}^{2}\; .
\eea
The overall mass-scale is not determined from oscillation experiments alone, but cosmological considerations imply~\cite{{Reid:2009nq},{GonzalezGarcia:2010un}}
\be
\sum_{i}m_{\nu_{i}}~\lesssim~0.6~~{\text eV}\; .
\label{numcos}
\ee
This still leaves an ambiguity in the ordering of the mass eigenstates. If $\Delta m^{2}_{32}>0$ neutrinos are said to be in a \emph{normal} mass hierarchy (NH) and if $\Delta m^{2}_{32}<0$ they are said to be in an \emph{inverted} mass hierarchy (IH).
\par
Similar to quarks and the CKM matrix, the electroweak and mass eigenstates in the lepton sector are related by a mixing matrix.  The relevant angles in this Pontecorvo-Maki-Nakagawa-Sakata (PMNS) mixing matrix~\cite{pmns} are denoted by $\theta_{23}$ (related to atmospheric oscillations), $\theta_{13}$ (reactor oscillations), and $\theta_{12}$ (relevant to solar oscillations). They are currently measured to be in the intervals~\cite{{PDG},{pastexpt1},{Abe:2010hy},{Adamson:2011ig},{Apollonio:2002gd},{Akiri:2011zz},{Wang:2011tp},{Jeon:2011zz}}

\bea
37^{\circ}\lesssim~&\theta_{23}\lesssim56^{\circ}\;,\nn \\
0^{\circ}\lesssim~&\theta_{13}\lesssim13^{\circ}\;,\nn \\
30^{\circ}\lesssim~&\theta_{12}\lesssim38^{\circ} \; .
\label{nuangles}
\eea
There are very strong indications~\cite{Adamson:2011ig} that the atmospheric angle $\theta_{23}$ is close to maximal ($\pi/4$). Solar data~\cite{pastexpt1}  also favors a `Large Mixing Angle' (LMA) solution that indicates large $\theta_{12}$. These intervals roughly translate to a PMNS matrix of the form
\be
\Bigl\lvert\mathcal{U}^{\text{\tiny{exp.}}}_{\text{\tiny{PMNS}}}\Bigr\rvert~\sim~
\begin{pmatrix}
0.8-0.9 &0.5-0.6& 0-0.2 \\
0.3-0.6&0.3-0.7&0.6-0.8 \\
0.1-0.5&0.5-0.8& 0.6-0.8
\end{pmatrix}~~~.
\label{exppmns}
\ee
Some comments are in order. The above matrix is very close to a \emph{tribimaximal} mixing matrix which has vanishing $\theta_{13}$ and maximal $\theta_{12},\, \theta_{23}$. The question of CP-violation in the lepton sector, even in the simplest case of neutrinos being Dirac particles, is at present open. This is intimately tied to the question of a vanishing $\theta_{13}$, since in the event of a vanishing or extremely small value the relevant Jarlskog-invariant ($J_{\text{\tiny{CP}}}\sim$ area of the unitarity triangle $\propto\sin 2\theta_{13}$) is also zero.
\par
Though almost all the observations from atmospheric, reactor and solar neutrino experiments can be accommodated conveniently in a three active-neutrino framework, there are a few outliers that may suggest existence of additional \emph{sterile} neutrinos (i.e. neutrinos that are electroweak singlets). Let us briefly review them.
\par
LSND employed a liquid scintillation detector to search for $\bar{\nu}_\mu\rightarrow\bar{\nu}_e$ oscillations~\cite{Aguilar:2001ty} with a baseline of about $30~\text{m}$. The neutrino energies were such that $L/E\sim\,\mathcal{O}(1)\,\text{(Km/GeV)}$. An excess of about 88 events was observed in the $20\,\text{MeV} \lesssim E \lesssim 60 \,\text{MeV}$ range. A conventional oscillation interpretation of the data requires $\Delta m^2 \sim 0.1-10\, \text{eV}^2$ and is referred to as the LSND anomaly~\cite{Aguilar:2001ty}.
\par
MiniBooNE was designed to test the LSND anomaly. They searched for both $\nu_\mu\rightarrow \nu_e$ and $\bar{\nu}_\mu\rightarrow\bar{\nu}_e$ oscillations~\cite{AguilarArevalo:2010wv}. The $E$ ($\sim1\,\text{GeV}$) and $L$ ($541\,\text{m}$) were both much bigger than LSND, but $L/E$ was still kept to be about the same. No oscillations in neutrino mode was observed above $475$ MeV, but an unexplained $3\sigma$ excess of $\nu_e$ events was found below $475\,\text{MeV}$. Subsequently, a $2.8\sigma$ excess of $\bar{\nu}_e$ was observed in the range $475\, MeV\leq E \leq 1250\, MeV$ consistent with LSND. There are preliminary updates from the MiniBooNE collaboration for their anti-neutrino data~\cite{MiniBooNE_new_antineu} where there is an excess below $475\,\text{MeV}$ similar to $\nu$ data, but the $\bar{\nu}_{\mu}\rightarrow\bar{\nu}_e$ signal above $475\,\text{MeV}$ has now diminished from before~\cite{MiniBooNE_new_antineu}.
\par
Recently, a re-evaluation of the reactor anti-neutrino flux~\cite{Mention:2011rk} also indicates a deficit from the expected value. The data sets used for the evaluation came from various experiments with very short baselines ($L<100\,\text{m}$). The observed to expected flux ratio was found to be $0.943\pm0.023$ compared to a previous value of $0.979\pm0.029$.
\par
There was also a puzzling discrepancy between muon neutrino and anti-neutrino disappearance data from the MINOS experiment~\cite{{Adamson:2011ig}, {Adamson:2011fa}} where the best-fit for the $\bar{\nu}$ data exhibited a higher $\Delta \overline{m}^2$ and lower $\sin^2 2\bar{\theta}$ than the $\nu$ data. Since it is a survival probability being measured, a resolution of both $\nu$ and $\bar{\nu}$ data requires an effective CPT violation, say for instance through some non-standard matter interactions~\cite{NSI}. There is an updated measurement from the MINOS collaboration in the $\bar{\nu}$ mode giving $\Delta\overline{m}^2<3.37\times10^{-3}\,\text{eV}^2$ at $90\%\,\text{C.L.}$ assuming $\sin^2 2\bar{\theta}=1.0$~\cite{Adamson:2011ch}. The most current, preliminary best-fits give $|\Delta \overline{m}^2|=\left[2.62^{+0.31}_{-0.28}(\text{stat})\pm0.09(\text{syst.})\right]\times 10^{-3}\text{eV}^2$, $\sin^2 2\overline{\theta}=0.95^{+0.10}_{-0.11}(\text{stat.})\pm0.01(\text{syst.})$ for the new anti-neutrino data~\cite{Adamson:2011ch}. The previous discrepancy between $\nu$ and $\bar{\nu}$ data therefore seems to be almost resolved. Prior to these measurements by MINOS, the strongest constraints on $\bar{\nu}$ parameters were from a global fit dominated by Super-Kamiokande data that included both atmospheric $\nu_\mu$ and $\bar{\nu}_\mu$ results~~\cite{{Adamson:2011ch},{ref:maltoni}}. 
\par
There have also been indications for a non vanishing $\theta_{13}$ from various experiments~\cite{{MINOS_theta13}, {Abe:2011sj}, {new-DCHOOZ}}. A non-zero $\theta_{13}$ has far-reaching implications for CP-violation in the lepton sector and the consistency of other neutrino parameter fits.
\par
MINOS detected $\nu_e$  appearance at the far detector~\cite{MINOS_theta13}, compared to expected background, suggesting for $\delta_{\text{\tiny{CP}}}=0$
\bea
2\sin^2\theta_{23}\sin^2 2\theta_{13}&<&~0.12~~~~~~\text{(NH)}\;,\nn \\
2\sin^2\theta_{23}\sin^2 2\theta_{13}&<&~0.20~~~~~~~\text{(IH)}\; .
\eea
The best-fit points~\cite{MINOS_theta13} for $2\sin^2\theta_{23}\sin^2 2\theta_{13}$ are deduced to be $0.041^{+0.047}_{-0.031}$ for normal (NH) and $0.079^{+0.071}_{-0.053}$ for inverted hierarchies (IH). $|\Dmq_{32}|=(2.32^{+0.12}_{-0.08})\times10^{-3}\,\text{eV}^2$, $|\Dmq_{21}|=(7.59^{+0.19}_{-0.21})\times10^{-5}\,\text{eV}^2$, $\theta_{23}=0.785\pm0.001$ and $\theta_{12}=0.60\pm0.02$ have been assumed in the above fits.
\par
 The T2K experiment~\cite{Abe:2011sj} observed six $\nu_e$ events that pass all selection criteria at the far detector. This suggests, at $90\%$ C.L., again for $\delta_{\text{\tiny{CP}}}=0$
\bea
0.03<~&\sin^2 2\theta_{13}&<0.28~~~~~~~\text{(NH)}\; ,\nn \\
0.04<~&\sin^2 2\theta_{13}&<0.34~~~~~~~~\text{(IH)}\; .
\eea
The best-fit points~\cite{Abe:2011sj} for $\sin^2(2\theta_{13})$ are found to be $0.11^{+0.1}_{-0.06}$ (NH) and $0.14^{+0.11}_{-0.08}$ (IH). The above T2K limits and best-fit values are extracted for $\sin^22\theta_{23}=1.0$ and $\Dmq_{32}=2.4\times 10^{-3}\,\text{eV}^2$.
\par
Most recently, there is a preliminary result~\cite{new-DCHOOZ} from Double-CHOOZ, based on the first 100 days of data, showing at $68\%\,\text{C.L.}$
\be
\sin^2 2\theta_{13}=0.085\pm0.029\,(\text{stat.})\pm0.042\,(\text{syst.})\;.
\ee
This result is particularly interesting since, being a reactor neutrino disappearance measurement, it is independent of CP phases and the mass hierarchy.
\par
Some of the above short-baseline discrepancies may be interpreted as being due to the presence of extra electro-weak singlet neutrinos. With this in mind, in the next section we briefly discuss the viability of sterile neutrinos.
\end{section}

\begin{section}{Sterile Neutrinos}
From the invisible Z-decay width and LEP measurements, the number of active-neutrino species is constrained to~\cite{PDG}
\bea
N^{\text{\tiny{Z-width}}}_{\nu}&=&2.92\pm0.05\;,\nn\\
N^{\text{\tiny{LEP}}}_{\nu}~~~&=&2.984\pm0.008\;.
\eea
Also, as already mentioned, most of the current neutrino oscillation data can be accommodated in a three active-neutrino framework. The Troitzk~\cite{troitsk} and Mainz~\cite{mainz} experiments give a mass bound of about
\be
m_\nu<2.3~\text{eV}\; ,
\ee
by measuring the endpoint region of the tritium $\beta$-decay spectrum.
In cosmology, neutrinos play a significant role by effecting the expansion history and growth of primordial perturbations which lead to a tighter mass bound~\cite{{Reid:2009nq},{GonzalezGarcia:2010un}}
\be
m_\nu\lesssim0.6~\text{eV}\; ,
\ee
for three flavor mixing. The KATRIN experiment~\cite{KATRIN} is speculated to reach a sensitivity of $m_{\nu}<0.2\,\text{eV}$.
\par
All these nevertheless still leave open the possibility of other neutrino species that are singlets under the SM gauge groups and therefore ``sterile". Trying to resolve the LSND and MiniBooNE anomaly with data from solar and atmospheric neutrino measurements require, as we noted in the last section, $\Delta m^2_{\text{\tiny{sterile}}} \sim \mathcal{O}(1)\, \text{eV}^2$. A similar mass squared difference is also required to reconcile the reactor anti-neutrino flux deficit. There are also some very controversial indications from the Heidelberg-Moscow experiment~\cite{KlapdorKleingrothaus:2004wj} of detecting a neutrino mass $0.17\,\text{eV}<~m_{\beta\beta}$. 
\par
An analysis~\cite{Dodelson:2005tp}, a few years ago, combining data from cosmic microwave background (CMB), large scale structure (LSS) and Lyman-$\alpha$ constrained the mass of a fourth sterile neutrino to be $m_s<0.23\,\text{eV}$ assuming they are thermal. In the non-thermal case it was shown that the constraints are non-trivial in the mass-density plane, but still viable. This was re-emphasized in a study~\cite{Hamann:2010bk} that combined the WMAP 7-year data, BBN, small-scale CMB observations and measurement of the Hubble parameter from the Hubble space telescope. The study concluded that the current data set mildly favors extra radiation in the universe and derived constraints on the number and mass of possible sterile neutrino species. More recently, in~\cite{Hamann:2011ge}, the authors conclude that though sterile neutrinos are disfavored by hot dark matter limits in minimal $\Lambda\text{CDM}$, extending the standard cosmological framework to include additional relativistic degrees of freedom or a dark-energy equation of state parameter $w<-1$ can relax these constraints substantially. 
\par
Constraints were also put on active-sterile mixing by measuring neutral-current interactions at MINOS~\cite{Adamson:2011ku}. By measuring the depletion of the neutral current event rate at the far detector, a $90\%$ limit was placed on the fraction of active neutrinos that transition to a sterile neutrino (assuming $\theta_{13}$=0)~\cite{Adamson:2011ku}
\bea
f_s=\frac{P_{\nu_\mu\rightarrow\nu_s}}{1-P_{\nu_\mu\rightarrow\nu_\mu}}~<~0.22~(0.40)\; .
\label{fs}
\eea
The number in the bracket is for the assumption $\theta_{13}=11.5^{\circ}$ and $\delta_{CP}=\pi$. 
\par
Apart from experiments reviewed in the last section, we should also mention two other experiments - SAGE~\cite{SAGE}  and GALLEX~\cite{GALLEX}. They give for $\nu_e$ disappearance a measured to calculated ratio $R=0.86\pm0.05$ consistent with each other. If interpreted as due to an additional sterile state, this corresponds to a $\Delta m^2_{\text{\tiny{sterile}}}=2.24~\text{eV}^2$ and $\sin^2 2\theta_{ee}=0.50$~\cite{Giunti_Ga}. This conclusion has been weakened though by a recent analysis~\cite{Conrad:2011ce}, based on its consistency with KARMEN and LSND data.
\par
Motivated by all these indications, extensive studies have been performed on fitting the short-baseline neutrino discrepancies to various models with sterile neutrinos and checking their consistency with other experiments~\cite{{Sorel:2003hf},{Nelson:2010hz},{Kopp:2011qd},{Giunti:2011ht},{Barger:2011rc}}. 
\par
 The inclusion of additional sterile neutrinos to the three active ones adds more structure to the neutrino oscillation formalism, without changing the basic framework.``$3+s$" refers to the case of $3$ active and $s$ sterile neutrinos. The case of 3 active neutrinos in the standard model shall henceforth be denoted as $3\nu\text{SM}$. For `n' neutrinos (active and sterile) the total number of angles in the mixing matrix is $n(n-1)/2$. The angles that rotate sterile states to sterile states are not relevant for neutrino oscillations and hence this number can be trimmed to $3(n-2)$ angles. Similarly, the number of CP phases (Dirac) that could be present in the PMNS matrix is $(n-2)(n-1)/2$. The number of physical CP phases relevant to electroweak physics and oscillations is $2n-5$, after field re-definitions. For ``$\,3+1$" this gives 6 angles, 3 phases and for ``$\,3+2$" this gives 9 angles, 5 phases.
\end{section}

\begin{section}{Neutrino Oscillations and the Short-baseline Limit.}
The probability for a neutrino state $\alpha$ oscillating into a state $\beta$ in vacuum is given by
\bea
P(\nu_\alpha \rightarrow \nu_\beta)&=& \delta_{\alpha\beta} - 4\sum_{i>j} \Re (U^*_{\alpha i} U_{\beta i} U_{\alpha j} U^*_{\beta j})\sin^2 (\Delta m^2_{ij} \frac{L}{4E})+ 2\sum_{i>j} \Im (U^*_{\alpha i} U_{\beta i} U_{\alpha j} U^*_{\beta j}) \sin (\Delta m^2_{ij} \frac{L}{2E})\;,
\label{fullosc}
\eea
where $i,j$ denote the mass eigenstates and $\alpha,\beta$ include both active and sterile neutrino eigenstates. $\Re$ and $\Im$ stand for the real and imaginary parts respectively. We will also use the notation $\Delta_{ij}=\Delta m^2_{ij} L/4E$. 
\par
In most oscillation experiments (which are constructed with sensitivity to a particular $\Dmq$) one can simplify the above by taking a two neutrino limit. In this limit we get the familiar result
\bea
P^{2\nu}(\nu_\alpha \rightarrow \nu_\beta)= \left\{ \begin{array}{ll}
1-\sin^22\theta \sin^2 (1.27\, \Delta m^2(\mbox{eV}^2) \frac{L\mbox{(Km)}}{E\mbox{(GeV)}}) & ; \textrm{ $\alpha=\beta$}\;,\\ 
~~\sin^22\theta \sin^2 (1.27\, \Delta m^2(\mbox{eV}^2) \frac{L\mbox{(Km)}}{E\mbox{(GeV)}}) & ; \textrm{ $\alpha\neq\beta$\;.}
\end{array} \right.
\label{eq38}
\eea
\par
 We note a few well known properties. In the two neutrino limit $P^{2\nu}(\nu_\alpha \rightarrow \nu_\beta; U_{eff})=P^{2\nu}(\nu_\beta \rightarrow \nu_\alpha; U_{eff})$. In general $P(\nu_\alpha \rightarrow \nu_\beta; U)=P(\nu_\beta \rightarrow \nu_\alpha; U^*)$. Under the assumption of CPT, in addition to having the mass-squared differences same for both neutrinos and anti-neutrinos, we also have $P(\bar{\nu}_\alpha \rightarrow \bar{\nu}_\beta;U)=P(\nu_\beta \rightarrow \nu_\alpha;U)$. The above two results lead to the fact that $P(\bar{\nu}_\alpha \rightarrow \bar{\nu}_\beta;U)=P(\nu_\alpha \rightarrow \nu_\beta;U^*)$. Thus, note that it is the last term in Eq.\,(\ref{fullosc}) that distinguishes neutrinos and anti-neutrinos (when the number of families is greater than 2) for $\alpha \neq \beta$, indicating CP violation (CPV).
 \par
  It is also important to emphasize that for disappearance measurements or survival probability  ($\alpha=\beta$) the last term vanishes. Hence, survival probabilities in vacuum will not be effected by any CP phases and must be the same for both neutrinos and anti-neutrinos if CPT holds. The last term being an odd-function of $\Delta m^2$ is also sensitive to the mass hierarchy of the neutrino species in principle.
 \par
 We will be interested in two specific limits of Eq.\,(\ref{fullosc}). The first limit is the short-baseline (SBL) limit which is relevant approximately when $L\sim\mathcal{O}(1)\,\text{Km}$, $E\sim\mathcal{O}(1)\,\text{GeV}$ and consequently $L/E\sim\mathcal{O}(1)\,\text{(Km/GeV)}$. The other limit we would be interested in is the long-baseline (LBL) limit where  $L\gtrsim\mathcal{O}(10^2)\,\text{Km}$ and $L/E\gtrsim\mathcal{O}(10^2)\,\text{(Km/GeV)}$. An intermediate case where $L\sim\mathcal{O}(1)\,\text{Km}$, $E\sim\mathcal{O}(10^{-3})\,\text{GeV}$ and $L/E\sim\mathcal{O}(10^3)\,\text{(Km/GeV)}$ is often called medium-baseline (MBL). 
 
\begin{table}
  \begin{tabular}{cccccccc}
  \hline\hline
 Model &~~~~~$\Delta m^2_{41}(\text{eV}^2)$&~~~~~$|U_{e4}|$ &~~~~~~$|U_{\mu 4}|$ &
    ~~~~~$\Delta m^2_{51}(\text{eV}^2)$ &~~~~~$|U_{e5}|$ &~~~~~~$|U_{\mu 5}|$ &~~~~~
    $\delta / \pi$ \\
    \hline
    ``$\,3+2$"&
    0.47& ~~0.128 &~~~ 0.165 &
    0.87& ~~0.138 &~~~ 0.148 &~~~ 1.64 \\
    \hline\hline
     \end{tabular}
  \caption{Global
    best-fit points using SBL data for the ``$\,3+2$" case ~\cite{Kopp:2011qd}.}
  \label{globalfit}
\end{table}

 \par
 We are going to primarily focus on a ``$\,3+2$"  scenario as analyzed in the recent global fits~\cite{Kopp:2011qd}, with additional mass squared differences in the $\mathcal{O}(1)\,\text{eV}^2$ range. The comprehensive global fits, to SBL neutrino experiments, in~\cite{Kopp:2011qd} use appearance data from LSND \cite{Aguilar:2001ty}, MiniBooNE~\cite{AguilarArevalo:2010wv}, KARMEN~\cite{Armbruster:2002mp}, NOMAD~\cite{Astier:2003gs} along with disappearance data from Bugey~\cite{Declais:1994su}, CHOOZ~\cite{Apollonio:2002gd}, Palo Verde~\cite{Boehm:2001ik} and CDHS~\cite{CDHS}. The most recent analysis further includes full spectral data from SBL reactor experiments ROVNO~\cite{ROVNO}, Krasnoyarsk~\cite{Krasnoyarsk}, ILL~\cite{ILL} and G\"{o}sgen~\cite{Gsgen} through rate measurements as summarized in~\cite{Mention:2011rk}. The main conclusion in the study is that the global fits to SBL oscillations, for a ```$\,3+2$" case, improves significantly with the inclusion of the new reactor anti-neutrino flux data, though some tension remains in the overall fit~\cite{Kopp:2011qd}. The global-fit values of~\cite{Kopp:2011qd} from SBL experiments is shown in Table \ref{globalfit}. We will adopt these values to analyze MBL/LBL measurements. Rather than taking the numbers in Table \ref{globalfit} as numbers set-in-stone, our attitude rather will be to view them as quantifying, to good extent, potential effects of sterile neutrinos in SBL.
 
\par
The SBL limit is the case most suitable for analyzing the LSND and MiniBoone experiments since they both were designed with a characteristic $L/E\sim 1\,\text{(Km)/(GeV)}$. A short-baseline (SBL) assumption leads to a simplification of the most general oscillation formula in Eq.\,(\ref{fullosc}). The following approximations may be made
\bea
\Delta m^{2}_{32}&\rightarrow&0\;, \nn \\ 
\Delta m^{2}_{21}&\rightarrow&0\; , 
\eea
compared to $\Delta m^{2}_{j1}$ where $j>3$. This is partly motivated by requirements from LSND and MiniBoone observations which require $\Delta m^{2}_{\text{\tiny{sterile}}}\sim \mathcal{O}(1)\,\text{eV}^2$. If there were no sterile neutrinos with $\Delta m^{2}_{\text{\tiny{sterile}}}\sim \mathcal{O}(1)\,\text{eV}^2$, then under the SBL approximation $P(\nu_\alpha\rightarrow\nu_\alpha; U)=1$, as should be expected since the baseline is not sufficient for significant oscillations into other flavors.
\par
Using the above assumptions in Eq.\,(\ref{fullosc}) we get for the ``$\,3+2$" case
\be
P_{\alpha\beta}(\text{CPV})_{\text{\tiny{SBL}}} =\delta_{\alpha\beta}-\sum_{i=4,5}4\Re(\chi^i_{\alpha\beta}) \sin^2 \Delta_{i1}-4\Re(\xi_{\alpha\beta})\sin^2 \Delta_{54}+\sum_{i=4,5}2\Im(\chi^i_{\alpha\beta}) \sin 2\Delta_{i1}+2\Im(\xi_{\alpha\beta})\sin 2\Delta_{54} \; ,
 \label{SBLeq}
\ee
where
\bea
\chi^i_{\alpha\beta}&=&(\delta_{\alpha\beta}-U_{\alpha4} U^*_{\beta4}-U_{\alpha5} U^*_{\beta5}) (U^*_{\alpha i} U_{\beta i})\; ,\nonumber \\ 
\xi_{\alpha\beta}&=& (U^*_{\alpha 5} U_{\beta 5} U_{\alpha 4} U^*_{\beta 4}) \; .
\eea
\par
For $\alpha=\beta=e$ this gives,
 \bea
\overline{P}(\bar{\nu}_e \rightarrow \bar{\nu}_e)_{\text{\tiny{SBL}}} &&= 1 - 4(1-|U_{e4}|^2-|U_{e5}|^2)\Big[|U_{e 4}|^2 \sin^2 \Delta_{41}+ |U_{e 5}|^2 \sin^2 \Delta_{51} \Big]-4|U_{e 5}|^2 |U_{e 4}|^2 \sin^2 \Delta_{54} \; . 
\eea

\begin{figure}
\begin{center}
\includegraphics[width=9.75cm,angle=0]{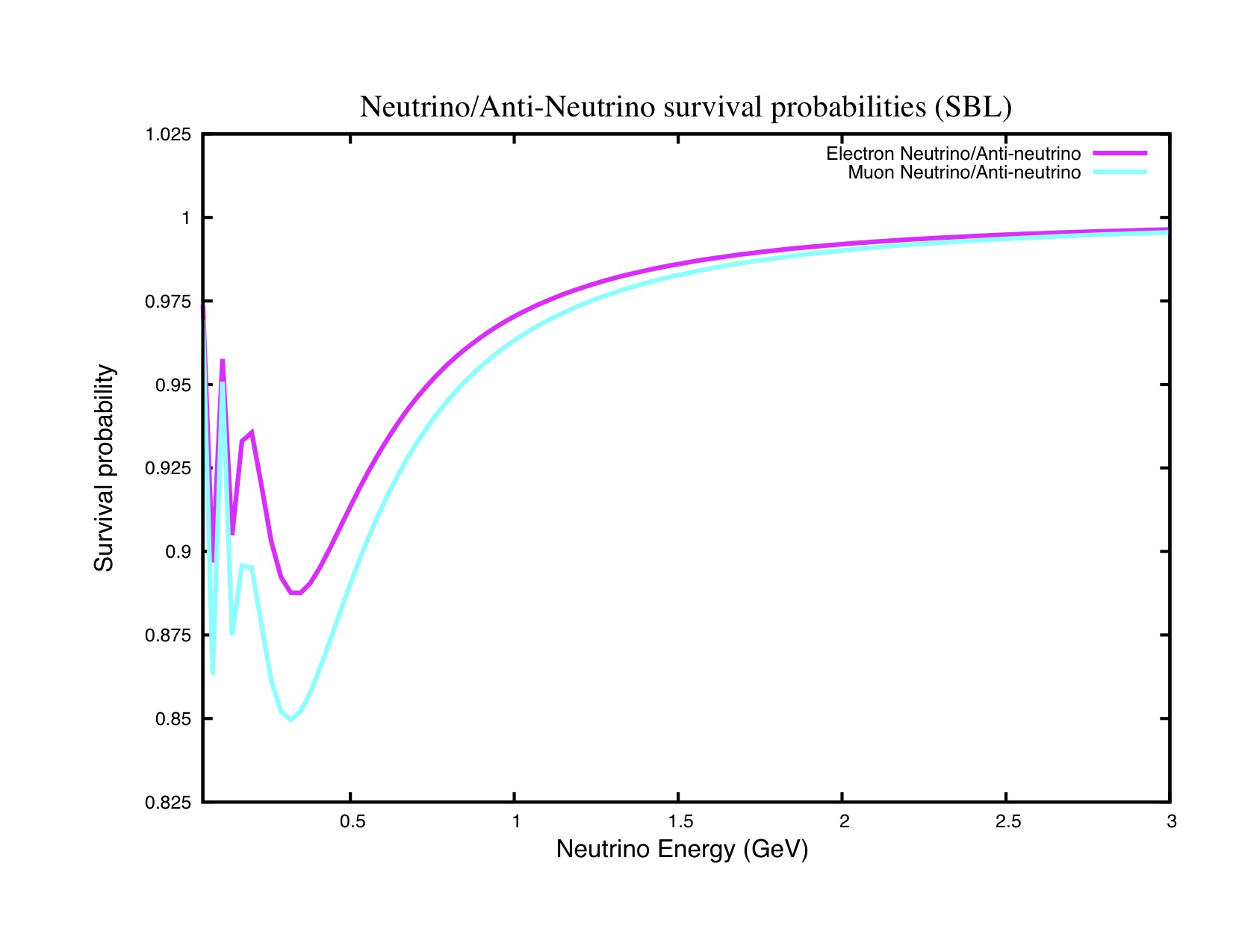}
\includegraphics[width=9.75cm,angle=0]{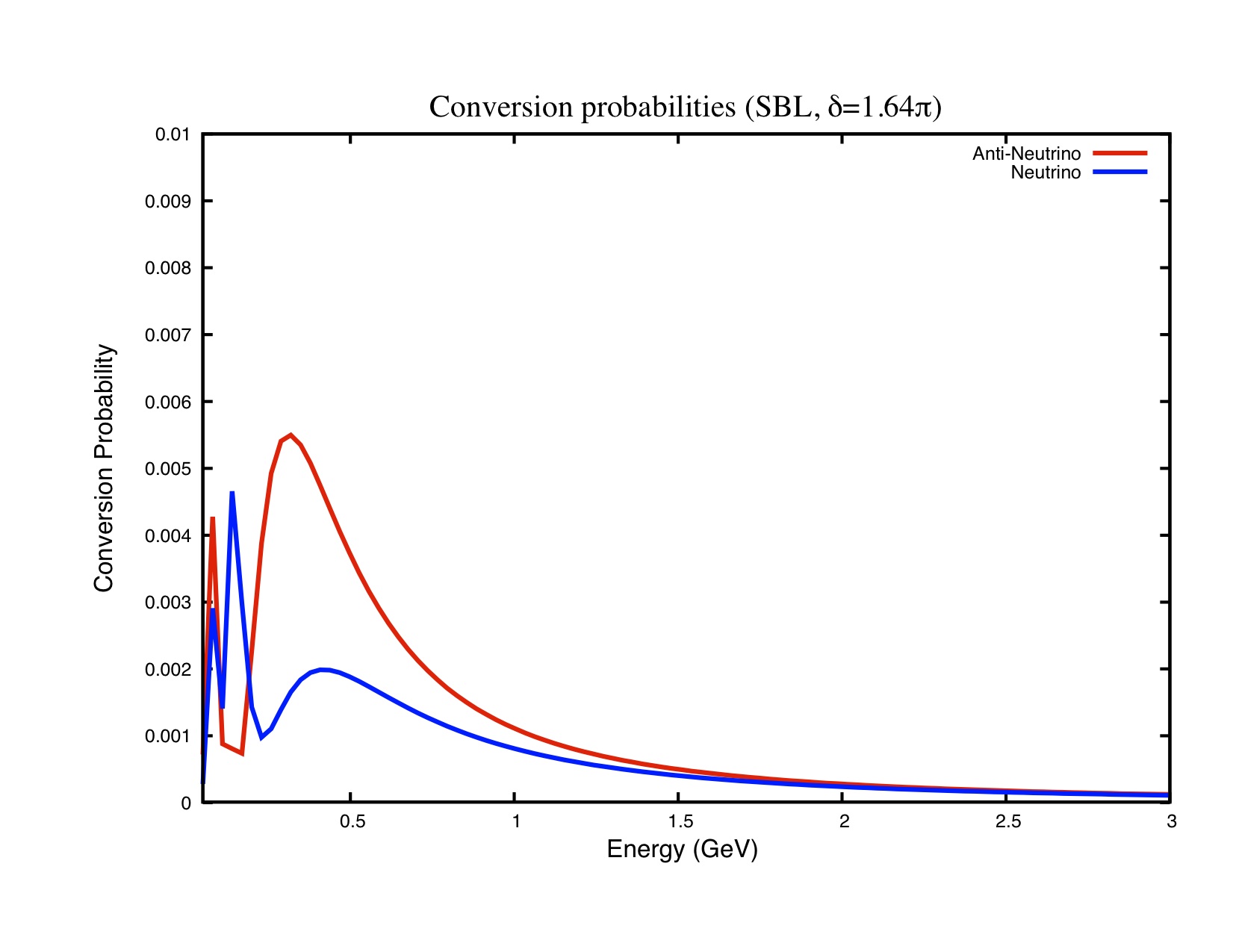}
\end{center}
\caption{Survival ($P_{ee}$ ) and conversion ($P_{\mu e}$) probabilities for neutrinos and anti-neutrinos in a SBL experiment with $L= 541\,\text{m}$, assuming $\delta=1.64\pi$ and parameters from Table \ref{globalfit}. The assumed distance to the detector corresponds to the baseline in MiniBooNE~\cite{AguilarArevalo:2010wv}. The difference in neutrino and anti-neutrino conversion probabilities is clearly visible in the bottom figure.}
\label{conversion_SBL}
\end{figure}

\par
When $\alpha \neq \beta$, the CP phase(s) can play a major role and the appearance or conversion probabilities may be different for neutrinos and anti-neutrinos. It may be shown that in the SBL approximation for ``$\,3+2$"there is only one relevant phase~\cite{Kopp:2011qd}. Consider the case $\alpha=\mu$ and $\beta=e$. Define the relevant phase as (also see Appendix A)
\be
\delta=\arg(U^*_{e 4} U_{\mu 4} U_{e 5} U^*_{\mu 5} ) \; .
\label{15phase}
\ee
Then
\bea
\Re(\xi_{\mu e})= |U^*_{e 4} U_{\mu 4} U_{e 5} U^*_{\mu 5} |\cos\delta \, ,~~~~~~~~~~~~~~~&&\Im(\xi_{\mu e})= |U^*_{e 4} U_{\mu 4} U_{e 5} U^*_{\mu 5} |\sin\delta\; ,\nn \\ 
\Re(\chi^{4}_{ \mu e})=-|U_{e4}|^2|U_{\mu 4}|^2-|U^*_{e 4} U_{\mu 4} U_{e 5} U^*_{\mu 5} | \cos\delta\, ,~&&\Im(\chi^{4}_{\mu e })=|U^*_{e 4} U_{\mu 4} U_{e 5} U^*_{\mu 5} | \sin\delta\; ,\nn \\ 
\Re(\chi^{5}_{\mu e})=-|U_{e 5}|^2|U_{\mu 5}|^2-|U^*_{e 4} U_{\mu 4} U_{e 5} U^*_{\mu 5} | \cos\delta\, ,~&&\Im(\chi^{5}_{\mu e})=-|U^*_{e 4} U_{\mu 4} U_{e 5} U^*_{\mu 5} | \sin\delta \;. 
\eea
As an aside, it should be mentioned that in the corresponding ``$\,3+1$" case there is no CP phase that is relevant to the SBL limit and therefore the conversion probabilities in that case cannot distinguish between $\nu$ and $\bar{\nu}$, for instance, at MiniBooNE. 
\par
With the inclusion of a CP phase, from Eq.\,(\ref{SBLeq}), we have for the conversion probabilities
\bea
P(\nu_\mu \rightarrow \nu_e; \text{CPV})_{\text{\tiny{SBL}}}=P(\nu_\mu \rightarrow \nu_e; \delta=0)_{\text{\tiny{SBL}}}+8|\xi_{\mu e}|\sin\Delta_{41} \sin\Delta_{51}\Big[\cos (\Delta_{54}-\delta)-\cos\Delta_{54}\Big]\; ,
\eea
\bea
\overline{P}(\bar{\nu}_\mu \rightarrow \bar{\nu}_e; \text{CPV})_{\text{\tiny{SBL}}}=P(\nu_\mu \rightarrow \nu_e; \delta=0)_{\text{\tiny{SBL}}}+8|\xi_{\mu e}| \sin \Delta_{41} \sin \Delta_{51}\Big[\cos (\Delta_{54}+\delta)-\cos \Delta_{54}\Big]\; .
\eea
\par
To give a more holistic picture we plot the disappearance and appearance probabilities in Fig.\,\ref{conversion_SBL}, assuming MiniBooNE base length, with the matrix elements and CPV phase of Table \ref{globalfit}. We see that for low energies the depletion of $\nu_{\mu}$ is greater than $\nu_{e}$ in the survival probability. Also note the expected enhancements and attenuations in the conversion probabilities, in this energy range, for $\bar{\nu}$ and $\nu$. The presence of a non zero CP phase could therefore, in principle, explain the difference between $\nu$ and $\bar{\nu}$ measurements in MiniBooNE~\cite{Karagiorgi:2006jf}. 
\par
As we shall see, the effects of sterile states at the near detector (ND) will have implications also in LBL measurements since usually it is a ratio between the far detector (FD) and ND neutrino fluxes that are compared, to probe for oscillations. Thus a depletion of the neutrino flux at the ND due to sterile states will affect MBL/LBL measurements too.
\end{section}


\begin{section}{Long-baseline Limit of Neutrino Oscillations.}
We now turn our attention to LBL experiments and it is our intention to understand in more detail the consequences of the SBL fits, or equivalently the presence of the two sterile neutrino states, to these experiments. Specifically, we explore in this section the impact of the SBL global fits to MINOS disappearance measurements and the recent $\theta_{13}$ determinations. The questions we would like to address are whether MINOS disappearance data can constraint or discriminate in a definite way the ``$\,3+2$" case from $3\nu\text{SM}$ and if the recent indications of a significant, non-vanishing reactor angle $\theta_{13}$ are effected drastically by sterile neutrino states, if they exist.

\par
In the LBL limit the assumption is that $L/E\,\text{(Km/GeV)}$ may be anywhere in the range $10^2-10^4$ or higher. The larger values are more appropriate for atmospheric and solar oscillations. For very large $\Delta m^{2}$, the oscillations get averaged out inside the detector and we may replace $\sin^2 (\Delta m^2 \frac{L}{4E}) $ by its expectation value ($1/2$). Consequently, $\Delta m^{2}_{41}$ and  $\Delta m^{2}_{51}$ are now averaged out since they are $\mathcal{O}(1)\,\text{eV}^2$.  $\Delta m^{2}_{21}$ may also be set to zero to first approximation. We will include this to be non-zero later, when discussing $\theta_{31}$ determination. 
\par
Calculating the oscillation probability from  Eq.\,(\ref{fullosc}), in the above limit, for ``$\,3+2$" gives
\bea
P_{\alpha\beta}(\text{CPV})_{\text{\tiny{LBL}}} &=&\delta_{\alpha\beta}-2\Re(\chi_{\alpha\beta})-4\Re(\xi_{\alpha\beta}) \sin^2\Delta_{54}-4\Re(\zeta_{\alpha\beta})\sin^2\Delta_{32}+\Im(\chi_{\alpha\beta})+2\Im(\xi_{\alpha\beta}) \sin2\Delta_{54}\nn \\
&+&2\Im(\zeta_{\alpha\beta})\sin2\Delta_{32}\; ,
 \label{lbl_surv}
\eea
where
\bea
\chi_{\alpha\beta}&=&(\delta_{\alpha\beta}-U_{\alpha4} U^*_{\beta4}-U_{\alpha5} U^*_{\beta5}) \sum_{i=4,5} U^*_{\alpha i} U_{\beta i} =\sum_{i=4,5}\chi^{i}_{\alpha\beta}\; ,\nn \\
\xi_{\alpha\beta}&=& U^*_{\alpha 5} U_{\beta 5} U_{\alpha 4} U^*_{\beta 4}\; , \nn \\
\zeta_{\alpha\beta}&=&(\delta_{\alpha\beta}-U_{\alpha3} U^*_{\beta3}-U_{\alpha4} U^*_{\beta4}-U_{\alpha5} U^*_{\beta5}) (U^*_{\alpha 3} U_{\beta 3})\; . 
\eea
\par
For a survival probability ($\alpha=\beta$) this may be re-written as
\be
P^{\text{\tiny{LBL}}}_{\alpha\alpha} \simeq(1-2\Re[\chi_{\alpha\alpha}])\left[1-\frac{4\Re[\zeta_{\alpha\alpha}]}{1-2\Re[\chi_{\alpha\alpha}]}\sin^2\Delta_{32}\right].
\label{surv_lbl_genform}
\ee
Here we have neglected the term proportional to $|\xi_{\alpha\beta}|$ which tends to be negligible numerically, since it is quartic in the sterile neutrino matrix elements. 
\par
We observe from Eq.\,(\ref{surv_lbl_genform}) that in the case of additional sterile states there is an effective \emph{normalization} factor modifying the survival probability as well as a \emph{modification} of the coefficient ($\sim\sin^22\theta$ in $3\nu\text{SM}$) of the energy dependent term. Thus, we conclude that the LBL survival probability $P^{\text{\tiny{LBL}}}_{\alpha\alpha}$ in the presence of sterile neutrinos is of a general form
\bea
P_{\alpha\alpha}^{\text{\tiny{LBL}}}\simeq\mathcal{N}_\alpha\left[ 1-\sin^2 2\vartheta^{\text{eff.}}_\alpha \sin^2\Delta_{32}  \right],~~~~~~
\label{survgenformLBL}
\eea
with
\bea
\mathcal{N}_\alpha&=&1-2\chi_{\alpha\alpha} \; ,\nn \\
\sin^2 2\vartheta^{\text{eff.}}_\alpha &=&\frac{4\,\zeta_{\alpha\alpha}}{1-2\chi_{\alpha\alpha}}\;.
\eea
The normalization factor ($\mathcal{N}_\alpha$), in principle, may be extracted by looking at asymptotically large neutrino energies. The $\sin^2 2\vartheta^{\text{eff.}}_\alpha$ coefficient determines the dip of the first minima. The modification to $\sin^2 2\vartheta$, when there are sterile neutrinos, has two parts - a direct modification of $\zeta_{\alpha\alpha}$ and a further scaling of this quantity by $\mathcal{N}_\alpha$. Observe that the \emph{quantities appearing in the LBL limit, $\mathcal{N}_\alpha$ and the modifying terms in $\sin^2 2\vartheta^{\text{eff.}}_\alpha$, are completely determined by SBL measurements through the matrix elements}.
\par
Using the global fit values from Table \ref{globalfit}, the various quantities that appear in the survival probability, Eq.\,(\ref{lbl_surv}), may be computed as (for an assumed $U_{e3}=0.1$ and $U_{\mu 3}=0.707$)
\bea
\chi_{ee}&=& 0.0342~~~~(3\nu \text{SM}:0) \;,\nn\\
\chi_{\mu\mu}&=& 0.0467~~~~(3\nu \text{SM}:0)\; ,\nn\\
\zeta_{ee}&=&0.0095~~~~(3\nu \text{SM}:0.0099)\; ,\nn\\
\zeta_{\mu\mu}&=&0.225~~~~~~(3\nu \text{SM}:0.25)\; .
\label{compvalues}
\eea
The numbers in the brackets are the corresponding values in the $3\nu\text{SM}$ case. Using the above values we get 
\bea
P_{ee}^{3+2} &=&0.932\left[ 1-0.0408 \sin^2 \Delta_{32} \right]~~\; ,~~\nn \\
P_{\mu\mu}^{3+2}&=&0.907\left[ 1-0.993 \sin^2 \Delta_{32}\right]~~\; ,~~
\label{FD_LBL_surv}
\eea
which is relevant to any FD in an MBL/LBL experiment, for instance, the MINOS FD. For comparison, in the $3\nu \text{SM}$ the expressions corresponding to above would have been
\bea
P^{3\nu\text{SM}}_{ee}&=&1-0.0396 \sin^2\Delta_{32}\; , \nn \\
P^{3\nu\text{SM}}_{\mu\mu}&=&1-\sin^2\Delta_{32} \; .
\eea
\par
Compared to the $3\nu \text{SM}$ prediction the overall flux is reduced by about $7\%$ for $\nu_e$ and by as large as $10\%$ for $\nu_\mu$ in Eq.\,(\ref{FD_LBL_surv}). The effective angle is seen to be not modified significantly in the survival probability and may still, in principle, be extracted to yield a value that is close to the true value. We will come back to this point again while discussing $|U_{\mu3}|$ and $|U_{e3}|$ determination. From Eqs. (\ref{compvalues}) and (\ref{FD_LBL_surv}) we note explicitly that the modification to $\sin^2 2\vartheta$, when there are sterile neutrinos, is due to a direct modification of $\zeta_{\alpha\alpha}$ and a further scaling of this quantity by $\mathcal{N}_\alpha$.
\par
It is also to be re-emphasized that when the sterile neutrino $\Dmq_{\text{\tiny{sterile}}}$ is large there will also generally be an effect in the ND, apart from the above effects in the FD, especially for low $E$. Since neutrino experiments usually compare a ratio of the fluxes at the FD and ND, taking into account geometric and other effects, modifications at the ND due to sterile states may also become relevant along with the FD effects. We will in fact see that for the values in Table \ref{globalfit}, the ND effects at MINOS are not completely negligible. 
\par
Although we specifically derived the expressions and numerical values for the ``$\,3+2$" case, \emph{a similar parametrization should be valid for any ``$\,3+s$" scenario} with $\Dmq_{\text{\tiny{sterile}}}\sim\mathcal{O}(1)\,\text{eV}^2$. Note also that owing to Eq.\,(\ref{survgenformLBL}) being a survival probability, any CP phases that may be present are completely irrelevant. This is particularly important since the CP phases that are relevant in LBL experiments may in general be mutually exclusive to the ones that are relevant in SBL experiments~\cite{Kopp:2011qd}. Thus, the above features must be applicable independently of CP phase structures in the LBL limit. The CP phases relevant to LBL may nevertheless become important in neutrino appearance measurements as we shall see.

\begin{subsection}{Sterile neutrinos and the MINOS disappearance data.}
\par
We now turn our attention to the analysis of  MINOS muon neutrino and anti-neutrino disappearance data~\cite{{Adamson:2011ig}, {{Adamson:2011fa},{Adamson:2011ch}}}. It is interesting to ask if the MINOS disappearance data can discriminate or put constraints on scenarios with sterile neutrinos, specifically ``$\,3+2$", as motivated by the SBL global fits. The quantity of interest here is the ratio of the observed events at the MINOS far detector ($L^{\text{\tiny{MINOS}}}_{\text{\tiny{FD}}}=734\,\text{Km}$, $4.2\,\text{Kt}$ fiducial mass) to that expected at the far detector if there were no neutrino oscillations. The latter is extrapolated from the MINOS near detector ($L^{\text{\tiny{MINOS}}}_{\text{\tiny{ND}}}=1.04\,\text{Km}$, $23.7\,\text{t}$ fiducial mass) through a Monte-Carlo, taking into account geometric and pion kinematic effects~\cite{{Adamson:2011ig}, {Adamson:2011fa},{Adamson:2011ch}}.
\par
Simple fits to the MINOS neutrino ($7.25\times 10^{20}$ protons on target (POT)) and anti-neutrino ($2.95\times 10^{20}$ POT) data~\cite{{Adamson:2011ig}, {Adamson:2011fa},{Adamson:2011ch}} are shown in Fig.\,\ref{neu_mu_MINOS} with various parametrizations, motivated by Eq.\,(\ref{survgenformLBL}). For the anti-neutrino analysis we have taken the new preliminary data~\cite{Adamson:2011ch}. 

\begin{figure}
\begin{center}
\includegraphics[width=14.0cm,angle=0]{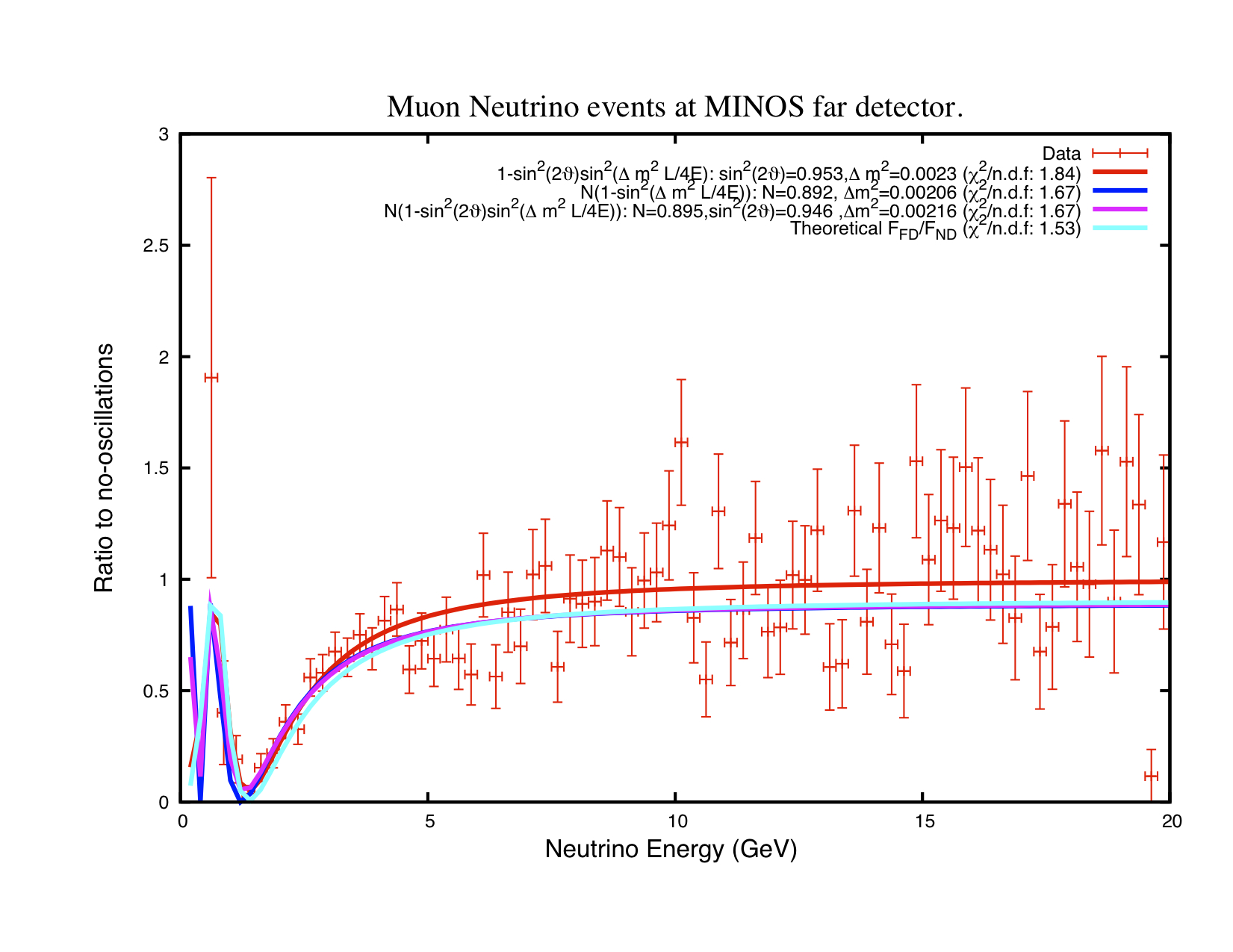}
\includegraphics[width=14.0cm,angle=0]{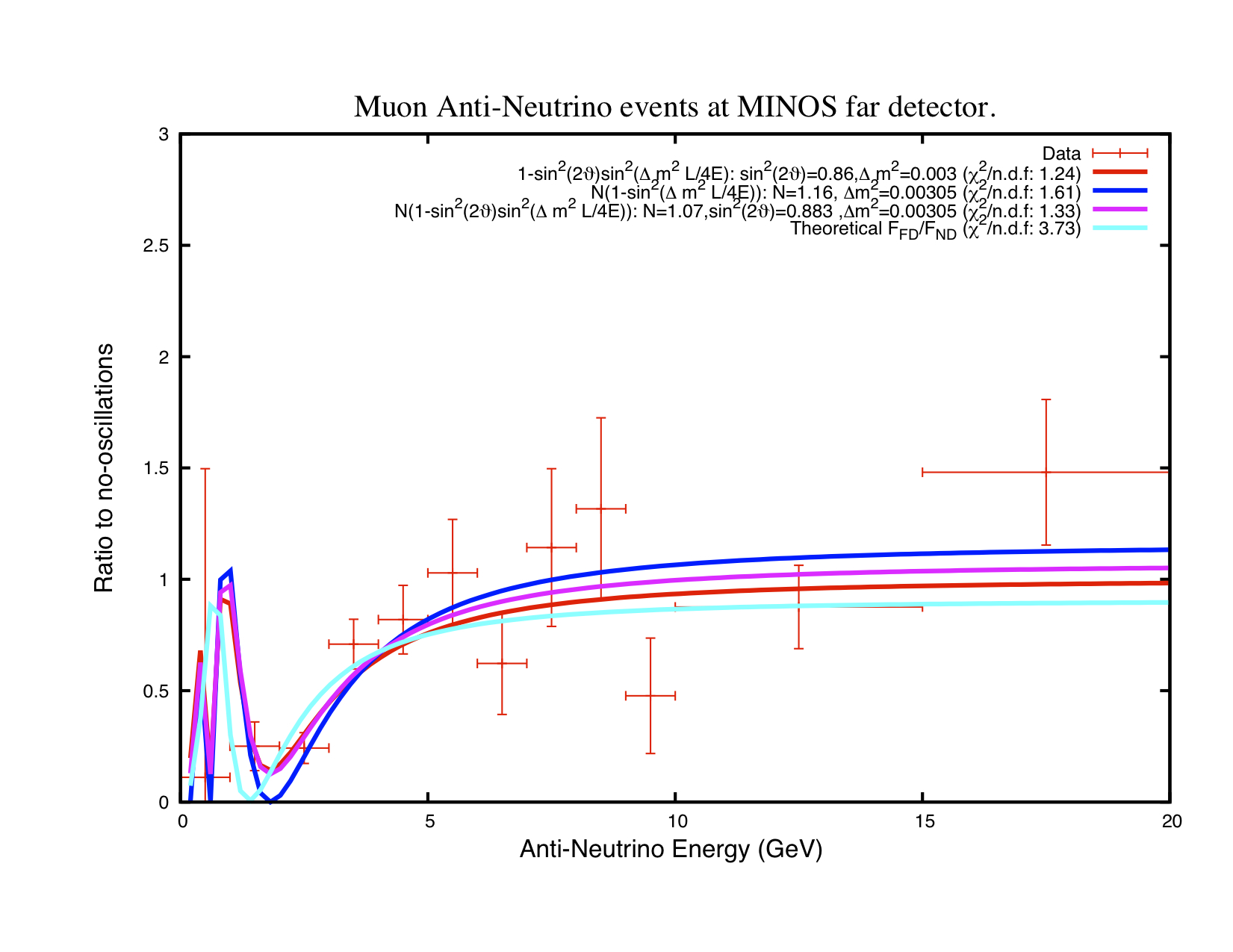}
\end{center}
\caption{Fits to MINOS neutrino~\cite{Adamson:2011ig} and anti-neutrino~\cite{{Adamson:2011fa},{Adamson:2011ch}} far-detector data assuming three different parametrizations motivated by Eq.\,(\ref{survgenformLBL}) - Case I (red), Case II (dark blue) and Case III (purple). The theoretical curve from Eq.\,(\ref{fdndMINOSratio}) (light blue), calculated based on the SBL gobal-fit values, is also shown. Though the plots are shown only till 20 GeV, the fits were done for the full range of neutrino energies in the data set ($50\,\text{GeV}$ for both $\nu_\mu$ and $\bar{\nu}_{\mu}$). The neutrino data corresponds to $7.25\times 10^{20}$ POT and the anti-neutrino data to $2.95\times 10^{20}$ POT.  }
\label{neu_mu_MINOS}
\end{figure}

\par
The values for the fit parameters are shown in Table \ref{MINOS_BestFits} along with their $1\sigma$ errors. In the first parametrization (Case I) both $\Delta m^2$  and $\sin^2 2\vartheta_\mu$ are floated, as in the analysis of the MINOS collaboration. A physical boundary on $\sin^2 2\vartheta$ is not imposed in the fit. In the second parametrization (Case II) $\Delta m^2$ and the normalization $\mathcal{N}_\mu$ are free parameters with $\sin^2 2\vartheta_\mu$ fixed at 1. The implicit motivation here is that, as we saw previously, the $\sin^2 2\vartheta_\mu$ coefficient is modified only minimally in disappearance measurements even when sterile neutrinos are present. In the final parametrization (Case III) we float $\mathcal{N}_\mu$, $\sin^2 2\vartheta_\mu$  and $\Delta m^2$.  In this context, it is worth re-emphasizing that the MINOS data shown is actually the ratio of the FD and ND neutrino fluxes and there will be deviations from Eq.\,(\ref{survgenformLBL}), especially at low energies for the values given in Table \ref{globalfit}, due to ND effects. 
\begin{table}
\begin{tabular}{ccccc}
\hline
\hline
Case&~~~$\Delta m^2(10^{-3}\text{eV}^2)$&~~~$\sin^2 2\vartheta_\mu$&~~~$\mathcal{N}_\mu$&~~~$\chi^2/\text{n.d.f}$\\
\hline
I&$2.31\pm0.10$&$0.953\pm0.04$&$1^{\dagger}$&$1.65$\\
II&$2.07\pm0.09$&$1^{\dagger}$&$0.895\pm0.03$&$1.48$\\
III&$2.17\pm 0.13$&$0.946\pm0.048$&$0.897\pm0.03$&$1.48$\\
$\mathcal{R}_{\mu\mu}$ &$-$&$-$&$-$&$1.53$\\
\hline\hline
\end{tabular}
\vspace{0.5cm}

\begin{tabular}{ccccc}
\hline
\hline
Case&~~~$\Delta \overline{m}^2(10^{-3}\text{eV}^2)$&~~~$\sin^2 2\overline{\vartheta}_\mu$&~~~$\overline{\mathcal{N}}_\mu$&~~~$\chi^2/\text{n.d.f}$\\
\hline
I&$3.0\pm0.23$&$0.86\pm0.08$&$1^{\dagger}$&$1.24$\\
II &$3.05\pm0.2$&$1^{\dagger}$&$1.16\pm0.13$&$1.61$\\
III &$3.05\pm 0.22$&$0.883\pm0.086 $&$1.07\pm0.122$&$1.33$\\
$\mathcal{R}_{\bar{\mu}\bar{\mu}}$ &$-$&$-$&$-$&$3.73$\\
\hline\hline
\end{tabular}
\caption{Parametric fits to MINOS neutrino and ant-neutrino data~\cite{{Adamson:2011ig},{Adamson:2011fa},{Adamson:2011ch}} for various parametrizations that take into account the possibility of sterile neutrinos, motivated by Eq.\,(\ref{survgenformLBL}). The symbol $\dagger$ denotes that the particular parameter is not floated in the fit under consideration.}
\label{MINOS_BestFits}
\end{table}

\par
The theoretical ratio of the flux at the FD and ND may be estimated from Eqs. (\ref{SBLeq}) and (\ref{lbl_surv}) as
\bea
\mathcal{R}_{\mu\mu}=\frac{F_{\text{\tiny{FD}}}(\mu)}{F_{\text{\tiny{ND}}}(\mu)}\simeq\frac{1-2\chi_{\mu\mu}-4\zeta_{\mu\mu}\sin^2 (\Delta M^2_{32} \frac{L^{\text{\tiny{MINOS}}}_{\text{\tiny{FD}}}}{4E})}{1-\sum_{i>3}4\chi^i_{\mu\mu} \sin^2 (\Delta M^2_{i1} \frac{L^{\text{\tiny{MINOS}}}_{\text{\tiny{ND}}}}{4E}) }\;.
\label{fdndMINOSratio}
\eea
This ratio estimated from the SBL global-fit values is also plotted in Fig.\,\ref{neu_mu_MINOS}. $|\Delta m^2_{32}|=2.32\times 10^{-3}\,\text{eV}^2$ has been assumed to calculate the theoretical curve. In the MINOS data set, the ND flux has been extrapolated to the FD, assuming no oscillations, including effects from beam-line geometry and meson decay kinematics~\cite{{Adamson:2011ig}, {Adamson:2011fa},{Adamson:2011ch}}. $R_{\mu\mu}$ and  $R_{\bar{\mu}\bar{\mu}}$, which are obtained assuming naively a collimated neutrino beam, may be directly compared to the real data sets to good approximation due to this. To quantify how the ratio predictions compare to the data sets, we may calculate a reduced-$\chi^2$ ($\chi^2/\text{n.d.f}$). It is found that  the theoretical ratio prediction has a reduced-$\chi^2$ of $1.53$ and $3.73$ for the neutrino and anti-neutrino data sets respectively (Table \ref{MINOS_BestFits}).
\par
From the neutrino data we extract an overall normalization, $0.897\pm0.03$, that is close to the one theoretically calculated in Eq.\,(\ref{FD_LBL_surv}) from the SBL global fits, which had a central value of $0.907$. The extracted $\sin^2 2\vartheta_\mu$ ($0.946\pm0.048$) is slightly lower than that predicted from the SBL global-fit values ($0.993$), but still within $1\sigma$. The uncertainties are larger in the anti-neutrino data set and the number of data points is also smaller. For this case it is found that both the extracted normalization and $\sin^2 2\vartheta$ are significantly larger and  smaller respectively than that predicted from SBL fits. 
There is hence some potential tension of the global-fit values with anti-neutrino data. Note that since this is a survival probability, if CPT is satisfied the probabilities in vacuum should be the same for neutrinos and anti-neutrinos.

\begin{figure}
\begin{center}
\includegraphics[width=8.0cm,angle=0]{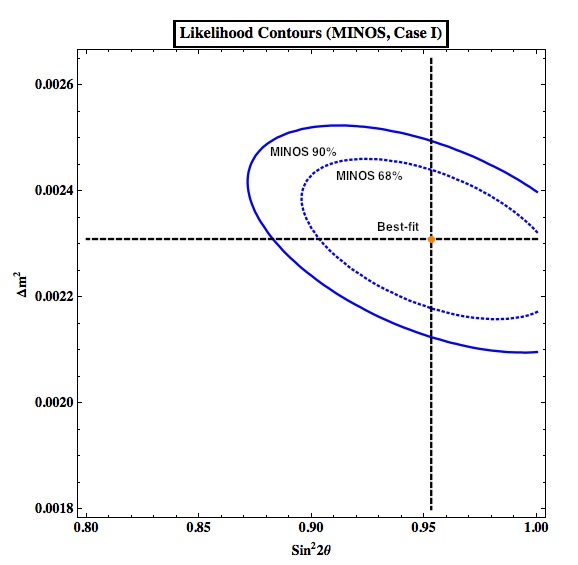}
\includegraphics[width=7.75cm,angle=0]{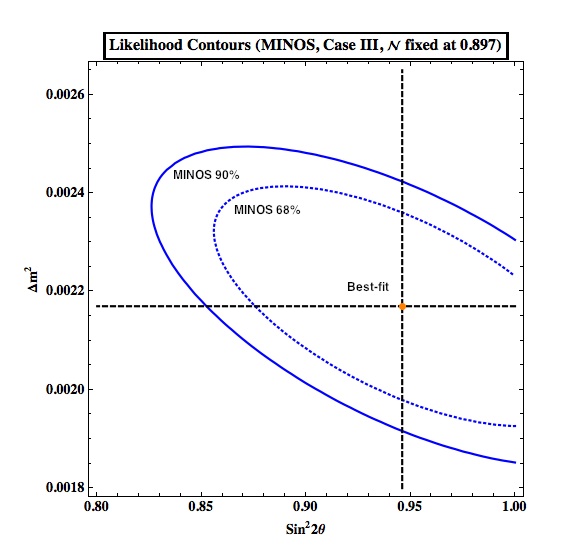}
\end{center}
\caption{Likelihood contours from our fitting procedure for MINOS neutrino data. The two plots are for Case I (left) and Case III (right) as defined in Table \ref{MINOS_BestFits}. Systematic errors have not been included. It is observed that our estimate of the $90\%\,\text{C.L.}$ bound for $\sin^2 2\vartheta$ shifts from $0.885$ to $0.853$, along with a reduction in $\Delta m^2$, when we include a normalization factor $\mathcal{N}_\mu$.}
\label{Likelihood_MINOS}
\end{figure}

\par
In Fig.\,\ref{Likelihood_MINOS} we plot the $68\%$ and $90\%$  likelihood contours for our simple fitting procedure, as applied to the MINOS neutrino data. Note that systematic uncertainties have not been taken into account as nuisance fit parameters. With the inclusion of a possible normalization $\mathcal{N}_\mu$ in the fits, the extracted $\Delta m^2$ decresases. The $90\%\,\text{C.L.}$ lower bound on $\sin^2 2\vartheta$ also shifts marginally to a smaller value ($0.90\rightarrow0.886$) as we include $\mathcal{N}_\mu$. Although we fix $\mathcal{N}_\mu$ in determining the $68\%$ and $90\%$ contours in Case III, the conclusions don't change significantly even if we include an error bar on $\mathcal{N}_\mu$.
 \par
 In the plots and fits shown, we have not changed the bin size from that given in the publicly available data set (the fits as done by the MINOS collaboration are for a bin size that is different from ours). We have nevertheless checked that re-binning the data (consistent with Fig. 2 of \cite{Adamson:2011ig}) does not significantly change our observations.  The MINOS collaboration also perform a more comprehensive likelihood analysis, compared to our fitting procedure, incorporating systematic uncertainties as nuisance parameters.
 For comparison, the values of the best-fits from the MINOS collaboration are $|\Delta m^2|=2.32^{+0.12}_{-0.08}\times 10^{-3}\,\text{eV}^2$, $\sin^2 2\vartheta>0.90\,(90\%\,\text{C.L.})$ for neutrinos~\cite{Adamson:2011ig} and $|\Delta \overline{m}^2|=\left[2.62^{+0.31}_{-0.28}(\text{stat})\pm0.09(\text{syst.})\right]\times 10^{-3}\text{eV}^2$, $\sin^2 2\overline{\vartheta}=0.95^{+0.10}_{-0.11}(\text{stat.})\pm0.01(\text{syst.})$ for the new anti-neutrino data~\cite{{Adamson:2011fa},{Adamson:2011ch}}.  So, although the central value is slightly different from ours, the lower bound on $\sin^22\vartheta$ is very close to the one obtained in our fit.  The comparison between  the $3\nu\text{SM}$ and ``$\,3+2$" cases shows only a small shift of this lower  bound and we expect this to be the case even after systematic errors are included.  
 Thus our simplified analysis and observations are still meaningful as long as we are comparing parameters and confidence levels extracted using the same fitting procedure, as in Table \ref{MINOS_BestFits} and Fig.\,\ref{Likelihood_MINOS}.
\par
Let us now try to derive some simple scaling relations between the $3\nu\text{SM}$ and ``$\,3+2$" confidence intervals. In general, from Eq.\,(\ref{surv_lbl_genform}), it may be shown that if $\alpha$ is a \emph{lower-bound} (say the $90\%\,\text{C.L.}$ limit) on the coefficient of $\sin^2\Delta_{32}$ in $P_{\mu\mu}$, the matrix element $|U_{\mu 3 }|$ would be constrained as 
\bea
 |U_{\mu3}|_{app.}^2\in\left[\frac{1-\sqrt{1-\alpha}}{2}~,~\frac{1+\sqrt{1-\alpha}}{2}\right]\; ,
\eea
when the `incorrect' assumption of $3\nu\text{SM}$ and no sterile neutrinos is made. With the assumption as ``$\,3+2$" we would have a constraint on the ``true" value
\bea
|U_{\mu3}|_{``true"}^2 \in\left[\frac{\beta'_{\mu}-\sqrt{{\beta'}_{\mu}^2-\mathcal{N}_{\mu}\alpha}}{2}~,~\frac{\beta'_{\mu}+\sqrt{{\beta'}_{\mu}^2-\mathcal{N}_{\mu}\alpha}}{2}\right] ,~~~~~
\label{musurv3p2gen}
\eea
where
\bea
\beta'_{\mu}&=&1-|U_{\mu 4}|^2-|U_{\mu 5}|^2\; ,\\
\mathcal{N}_{\mu}&=&1-2\chi_{\mu\mu}= 1-2\beta'_\mu\left(1-\beta'_\mu\right)\; .
\eea
It is easy to see from the above equations that the \emph{allowed interval for the extracted $|U_{\mu 3}|$ is shifted slightly, to lower values, when one includes sterile neutrinos in the extraction procedure}. Also note that in the case of the sterile-neutrinos, due to the modification, \textit{the coefficient of  $\sin^2{\Delta_{32}}$ (which would have been $\sim\sin^2 2\theta_{23}$ in the $3\nu\text{SM}$ case) can never be unity}. 
\par
For the $\sin^2 2\vartheta>0.885\,(90\%\,\text{C.L.})$ $3\nu\text{SM}$ (Case I) bound of Fig.\,\ref{Likelihood_MINOS} (left plot) this translates to
\bea
|U_{\mu3}|_{app.}^2 &\in& \left[0.33,0.67\right]\;,
\eea
and for the or the $\sin^2 2\vartheta>0.853\,(90\%\,\text{C.L.})$ ``$\,3+2$"  scenario (Case III), Fig.\,\ref{Likelihood_MINOS} (right plot), it becomes
\bea
|U_{\mu3}|_{``true"}^2 &\in& \left[0.29,0.66\right]\;.
\eea
The extracted $|U_{\mu 3 }|$ indeed shifts downwards by a few percent from the $3\nu\text{SM}$ case, when ``$\,3+2$" is assumed. This will be relevant to us when we discuss $\nu_e$ appearance measurements, since the value assumed for $|U_{\mu 3 }|$ will have a bearing on the extracted $|U_{e3}|$ in that case.

\par
It is clear from the analysis of the MINOS neutrino and anti-neutrino data sets that they by themselves cannot distinguish between the $3\nu\text{SM}$ and ``$\,3+2$" cases in a definite way. There nevertheless is a marginal improvement in the reduced-$\chi^2$, in the neutrino fits, when sterile neutrinos are included. The normalization as extracted from the neutrino data is close to that theoretically predicted for the FD and including any ND effects from sterile states improves the fit slightly. The uncertainties in the anti-neutrino data are larger and the reduced-$\chi^2$ obtained for the theoretical prediction is poor. With higher statistics the analysis in the anti-neutrino sector may be improved in the future.
\par
We will now explore the effects of sterile neutrinos on $\theta_{13}$ determination, in the context of $\nu_e$ disappearance and appearance measurements.
\end{subsection}

\begin{subsection}{Effects of sterile neutrinos on the determination of $\theta_{13}$}
There have been recent indications from experiments like T2K, MINOS and Double-CHOOZ for a possibly non-zero $\theta_{13}$. Let us try to understand the implications of sterile neutrinos for $\theta_{13}$ determination in these present and upcoming MBL/LBL experiments. If there are sterile neutrinos with $\Delta m_{\text{\tiny{sterile}}}^2\sim\mathcal{O}(1)\,\text{eV}^2$, then the mixing angles corresponding to them may be extracted in principle solely from SBL experiments. These matrix elements thus extracted, may then be used in the analysis of MBL/LBL experiments that aim to measure the reactor angle $\theta_{13}$. Due to the presence of the additional sterile states there may be possible modifications to the extracted $\theta_{13}$ angle or more precisely the extracted matrix element $|U_{e3}|$.

%

\subsubsection {\bf Reactor $\nu_e$ disappearance measurements}
\par
Let us first consider neutrino/anti-neutrino disappearance experiments which measure survival probabilities. The CHOOZ~\cite{Apollonio:2002gd} experiment and its successors Double-CHOOZ~\cite{Akiri:2011zz}, Daya Bay~\cite{Wang:2011tp} and RENO~\cite{Jeon:2011zz} are examples in this category. We take Double-CHOOZ as an example for purposes of our discussion. At Double-CHOOZ the FD is at $1050\,\text{m}$ and the ND is expected to be placed $400\,\text{m}$ from the reactor cores. For these base lengths and $E$ ($\sim\,3\,\text{MeV}$) we may use the MBL approximation at the FD.
\par
We have for the $\nu_e/\bar{\nu}_e$ survival probability, at the FD
\bea
P^{ee}_{\text{\tiny{LBL}}}&=& 1-2\left(|U_{e4}|^2+|U_{e5}|^2-|U_{e4}|^4-|U_{e5}|^4-|U_{e4}U_{e5}|^2\right)-4\left(|U_{e3}|^2(1-|U_{e3}|^2)-(1-\beta'_e) |U_{e3}|^2\right)\sin^2\Delta_{31}\nn\\
&-&4\left(|U_{e2}|^2(1-|U_{e2}|^2)-(1-\beta'_e) |U_{e2}|^2\right)\sin^2\Delta_{21}+8|U_{e2}|^2|U_{e3}|^2\sin\Delta_{21} \sin\Delta_{31}\cos\Delta_{32}\; ,
\eea
where, as before
\bea
\beta'_{e}&=&1-|U_{e 4}|^2-|U_{e 5}|^2\; .
\eea
Note that we have now retained $\Delta m_{21}^2$ terms explicitly. Since this is a survival probability, any CP phase that may be present in the MBL/LBL limit is completely irrelevant and does not cause any ambiguities. This is not the case, as we shall see, when we consider conversion probabilities where the phases may play a significant role. For the Double-CHOOZ baseline and characteristic $E$ ($\sim\,3\,\text{MeV}$), the terms which are proportional to $\Delta_{21}$ and quartic in sterile-neutrino matrix elements may be dropped without incurring significant errors. This leads to the familiar expression
\be
P^{ee}_{\text{\tiny{LBL}}}\simeq 1-2\chi_{ee}-4\zeta_{ee}\sin^2\Delta_{32}\; ,
\ee
derived in Eq.\,(\ref{surv_lbl_genform}) before. This may now be rewritten as
\be
P^{ee}_{\text{\tiny{LBL}}}\simeq \mathcal{N}_e\left[1-\sin^22 \vartheta_e\sin^2\Delta_{32}\right]\;,
\ee
where
\bea
\mathcal{N}_e&=& 1-2\chi_{ee}\simeq1-2(1-\beta'_e)\nn\;,\\
\sin^22 \vartheta_e&=&\frac{4\,\zeta_{ee}}{1-2\chi_{ee}}\simeq4\frac{|U_{e3}|^2(\beta'_e-|U_{e3}|^2)}{ 1-2(1-\beta'_e)}\;.
\label{edissparams}
\eea
\par
Under the 3-neutrino assumption this survival probability may have been written as
\be
P^{ee}_{\text{\tiny{LBL}}}\simeq1-4 |U_{e3}|_{app.}^2 (1-|U_{e 3}|_{app.}^2)\sin^2\Delta_{32} \;.
\label{3nuesurvchooz}
\ee
From this we may associate an ``apparent"  value for the $\theta_{13}$ angle, through the relation
\be
\sin^2 2\theta^{app.}_{13}= ~4|U_{e3}|_{app.}^2(1-|U_{e3}|_{app.}^2)\; ,
\ee
leading to the standard form for the $\nu_e\rightarrow\nu_e$ survival probability
\be
P^{ee}_{\text{\tiny{LBL}}}\simeq1-\sin^2 2\theta^{app.}_{13}\sin^2\Delta_{32} \;.
\ee
\par
In general, if $\alpha'$ is an \emph{upper-bound} on the coefficient of $\sin^2\Delta_{32}$ in the survival probability $P_{ee}$, we have the constraint
\bea
 |U_{e3}|_{app.}^2 ~\leq~\frac{1-\sqrt{1-\alpha'}}{2}\; ,
\eea
for the apparent value when the ``incorrect" assumption of no sterile neutrinos is made. With the correct assumption we would have a constraint on the true value 
\bea
|U_{e3}|_{true}^2\equiv|U_{e3}|^2 ~\leq~\frac{\beta'_e-\sqrt{{\beta'}_e^2-\mathcal{N}_e\alpha'}}{2}\;.
\label{surv3p2gen}
\eea
Compared to the $|U_{\mu 3}|$ extraction case in Eq.\,(\ref{musurv3p2gen}), we have dropped a solution that is already ruled out. Also note that in the probability expressions, leading to the above extraction, the $\mathcal{N}_e$ factors cancel while taking the FD/ND ratio. This has to do with the fact that both the FD and ND probability expressions are usually more akin to the MBL limit in reactor neutrino experiments, for instance for a ND around $400\,\rm{m}$ as in Double-CHOOZ.
\par
 In Table \ref{ue3dissvals} we list for comparison, values of $|U_{e3}|$ and $4\,|U_{e3}|^2(1-|U_{e3}|^2)$ determined in the neutrino disappearance case, using the $3\nu\text{SM}$ (``apparent") and ``$\,3+2$" (``true") assumptions. The true values for $|U_{e3}|$ are found to be generally smaller than the apparent values, as is to be expected from Eq.\,(\ref{surv3p2gen}). Nevertheless we observe that, due to the absence of CP phase ambiguities, negligible matter-effects and smallness of the sterile-neutrino matrix elements, the value of $|U_{e3}|$ extracted in reactor neutrino disappearance experiments are not modified significantly even when sterile neutrinos are present. Due to this, the extracted value of $\theta_{13}$ even under the incorrect assumptions is still close to the true value. 
\begin{table}
\begin{tabular}{cccc}
\hline
\hline
$|U_{e3}|_{\text{\tiny{app.}}}$&~~~$4\,|U_{e3}|^2_{\text{\tiny{app.}}}(1-|U_{e3}|^2_{\text{\tiny{app.}}})$&~~~$|U_{e3}|$&~~~$4\,|U_{e3}|^2(1-|U_{e3}|^2)$\\
\hline
0.224&~~0.19&~~0.219&~~0.183\\
0.198&~~0.15&~~0.194&~~0.145\\
0.168&~~0.11&~~0.165&~~0.106\\
0.133&~~0.07&~~0.131&~~0.067\\
0.087&~~0.03&~~0.085&~~0.029\\
\hline\hline 
\end{tabular}
\caption{Comparison of some representative apparent and true values of measured $|U_{e3}|$ and $4\,|U_{e3}|^2(1-|U_{e3}|^2)$ in a \emph{neutrino disappearance experiment}. A ``$\,3+2$" scenario is assumed and the matrix elements for the estimate are taken from the SBL global fits.}
\label{ue3dissvals}
\end{table}
\par
The above observations are especially pertinent in view of the recent preliminary result from Double-CHOOZ ~\cite{new-DCHOOZ}, suggesting
\be
\sin^2 2\theta^{\text{\tiny{D-CHOOZ}}}_{13}=0.085\pm0.029\,(\text{stat.})\pm0.042\,(\text{syst.})\;.
\ee
With a Double-CHOOZ FD at $1050\,\text{m}$ and a future ND at $400\,\text{m}$, the FD/ND ratio will be of the form $\sim\text{MBL}/\text{MBL}$ and as pointed out before Eq. (\ref{surv3p2gen}) the normalization factor $\mathcal{N}_e$ would cancel in the numerator and denominator in this case. Although Double-CHOOZ currently lacks a ND, they normalize their measurement to the Bugey experimental data~\cite{Declais:1994su}, which accounts for an approximate  0.945 suppression factor with respect to the expected one.  The result is then fitted to the form $1-\sin^22\theta_{13} \sin^2 \Delta_{32}$ to extract the value of $\sin^22\theta_{13}$. The difference between the normalization factor extracted from the Bugey experiment and the ${\mathcal N}_e$ predicted in the ``$\,3+2$" neutrino scenario is small compared to the systematic and statistical errors. We expect that this small difference, between the energy independent normalization factors, will not lead to any relevant variation of the extracted $\sin^22\vartheta$ value obtained from the fit. Of course if the full energy dependence for events at the Bugey detector at $15\,\text{m}$ is considered, one should use an expression $\mathcal{R}_{ee}$ analogous to $\mathcal{R}_{\mu\mu}$ used in Eq. (\ref{fdndMINOSratio}) for MINOS. 

Taking the central value above and assuming that there are two additional sterile states, characterized by the values in Table \ref{globalfit}, gives
\bea
|U_{e3}|^{\text{\tiny{D-CHOOZ}}}_{\text{\tiny{app.}}}&=& 0.147\,,~4|U_{e3}|^2_{\text{\tiny{app.}}}(1-|U_{e3}|^2_{\text{\tiny{app.}}})^{\text{\tiny{D-CHOOZ}}}=0.085\;,\nn\\
|U_{e3}|^{\text{\tiny{D-CHOOZ}}}&=&0.145\,,~4|U_{e3}|^2(1-|U_{e3}|^2)^{\text{\tiny{D-CHOOZ}}}=0.082\; .
\eea
As emphasized before, the difference between the extracted values in the 3-neutrino scenario and in the ``$\,3+2$" scenario is very small compared to the current errors.
\par
Now, consider a standard parametrization of the ``$\,3+2$" PMNS matrix 
\bea
\mathcal{U}^{3+2}_{\text{\tiny{PMNS}}}=\prod^{3}_{j>i,i=1}~\mathbb{R}_{ij}\; ,
\eea
where the product is to be done from right to left and $\mathbb{R}_{ij}$ is a complex or real rotation matrix in the $ij$-plane (see Appendix A). Let us generically denote by $\theta_s$ the small sterile angles ($\theta_{ij},\,j>3$). Using this parametrization, it may be seen that the coefficient in Eq.\,(\ref{edissparams}) is still very close to $\sin^2 2 \theta_{13}$, the deviations being of $\mathcal{O}(\theta^4_{s})$ (Appendix A). The quantity $|U_{e3}|$ is nothing but $\sin\theta_{13}$ in the $3\nu\text{SM}$ case. In the ``$\,3+2$" case, using the standard parametrization, it gets modified to $\cos\theta_{14}\cos\theta_{15}\sin\theta_{13}$ (see Appendix A). The quantity $4\,|U_{e3}|^2(1-|U_{e3}|^2)$ in the $3\nu\text{SM}$ case would have corresponded exactly to $\sin^2 2\theta_{13}$. In the ``$\,3+2$" case it deviates from $\sin^2 2 \theta_{13}$ by terms of $\mathcal{O}(\theta^2_s)$ (Appendix A). Note also that $J_{\text{\tiny{CP}}}\propto\,\sin2\theta_{13}$ in the $3\nu\text{SM}$ case.
\par
All the above conclusions are also applicable to upcoming experiments like Daya Bay~\cite{Wang:2011tp} and RENO~\cite{Jeon:2011zz} that aim to measure $\theta_{13}$ through a disappearance measurement. This is to be contrasted with appearance measurements that we discuss next.

\subsubsection{\bf $\nu_e$ appearance measurements}
\par
The value of $\theta_{13}$ may also be deduced by looking for $\nu_{e}$ appearance in experiments that measure neutrino conversion probabilities. This includes experiments such as T2K~\cite{Abe:2011sj} and MINOS~\cite{MINOS_theta13}. 
\par
Let us focus on T2K~\cite{Abe:2011sj} for the purposes of our discussion. At T2K the neutrino energy peaks around $0.6\,\text{GeV}$ and the ND(s) and FD are at $280\,\text{m}$ and $295\,\text{Km}$ respectively. This ensures $\Delta_{32}\sim\pi/2$ at the FD when $E\sim0.6\,\text{GeV}$, giving an oscillation maximum. These base lengths also enable us to use all the approximations for LBL again at the FD. The ND effects can again be quantified and studied as before.
\par
The transition probability in vacuum, keeping $\Delta m^2_{21}$ explicitly, may be written as
\bea
P_{\text{\tiny{LBL}}}^{\mu e}&=& 4|U_{\mu 3}|^2 |U_{e3}|^2\sin^2\Delta_{31}+4|U_{\mu 2}|^2 |U_{e2}|^2\sin^2\Delta_{21}+ 8 |U^{*}_{\mu 3}| |U_{e 3}| |U_{\mu 2}| |U^{*}_{e2}|    \sin\Delta_{31} \sin\Delta_{21}\cos (\Delta_{32}-\delta_3)\nn\\
&+&4 |U_{\mu 3}| |U_{e3}| |\beta''| \sin\Delta_{31}\sin (\Delta_{31}-\delta_1)+4 |U_{\mu 2}| |U_{e2}| |\beta''| \sin\Delta_{21}\sin (\Delta_{21}-\delta_2)+ 2\big(|U_{\mu 4}|^2  |U_{e 4}|^2+ |U_{\mu 5}|^2  |U_{e5}|^2\nn\\
&+& |U_{\mu 4}|  |U^*_{e4}|    |U^*_{\mu 5}|  |U_{e5}| \cos\delta\big) ~~\;,
\label{mueconversion}
\eea
where
\bea
\beta''&=&\sum_{i\geq4} U_{\mu i}^{*} U_{ei}\; , \nn \\
\delta_1 &=& \arg\left( U^*_{\mu 3} U_{e3} \beta'' \right) \;,\nn \\
\delta_2 &=& \arg\left( U^*_{\mu 2} U_{e2} \beta'' \right) \;,\nn \\     
\delta_3 &=& \arg\left( U^*_{\mu 3} U_{e3} U_{\mu 2} U^*_{e2}\right) \;.                                           
\eea

\par
The only approximation we have made in Eq.\,(\ref{mueconversion}) is to average terms containing large sterile mass-squared differences. In contrast to the previous case there now appears non-trivial CP phases, as this is a conversion probability. Due to the presence of these phases there could be interesting interferences between the various terms and it's seen that one can no longer drop terms, as we did in the survival probability case, without significant errors. 
\par
These phases which appear in the LBL limit are in general independent of the phase $\delta$ extracted from the SBL global fits. Also note that when $\beta''\neq0$ the phase $\delta_3$ is not independent and is given by $\delta_1-\delta_2$. The effective CP phases $\delta_1$ and $\delta_2$ may be related to the `fundamental' CP phases $\delta_{12}$ and $\delta_{13}$ in some particular parametrization of  $\mathcal{U}^{3+2}_{\text{\tiny{PMNS}}}$ (Appendix A).
\par
To get a better understanding of what the various terms in Eq.\,(\ref{mueconversion}) mean, we briefly look at the corresponding expression in the $3\nu\text{SM}$ case. In the $3\nu\text{SM}$ the conversion probability has the well-known form
\bea
P^{\text{\tiny{LBL}}}_{\mu e} \simeq  P^{3\nu\text{\tiny{SM}}}_{\text{\tiny{ATM.}}}+P^{3\nu\text{\tiny{SM}}}_{\odot}+2\sqrt{P^{3\nu\text{\tiny{SM}}}_{\odot} P^{3\nu\text{\tiny{SM}}}_{\text{\tiny{ATM.}}}}\cos(\Delta_{32}-\delta_3)\, ,
\label{3nsmconv}
\eea
with 
\bea
 P^{3\nu\text{\tiny{SM}}}_{\text{\tiny{ATM.}}}&\cong&\sin^2\theta_{23}\sin^2 2\theta_{13}\sin^2\Delta_{31}\; ,\nn\\
 P^{3\nu\text{\tiny{SM}}}_{\odot}&\cong&\cos^2\theta_{23}\sin^2 2\theta_{12}\sin^2\Delta_{21}\; .
\eea
\par
The first and second terms in Eq.\,(\ref{3nsmconv}) are the atmospheric and solar oscillation contributions. The last term denotes an `interference'  between the atmospheric and solar oscillations with a relative phase shift $\delta_3$. Label this term $P^{\text{\tiny{INT.}}-\delta_3}_{\odot-\text{\tiny{ATM.}}}$ and in terms of the matrix elements it is
\be
P^{\text{\tiny{INT.}}-\delta_3}_{\odot-\text{\tiny{ATM.}}}\equiv 8 |U^{*}_{\mu 3}| |U_{e 3}| |U_{\mu 2}| |U^{*}_{e2}|    \sin\Delta_{31} \sin\Delta_{21}\cos (\Delta_{32}-\delta_3)\;.
\ee
If we define $\delta_{CP}=-\arg(U_{e3})$, then for small $|U_{e3}|$ values, $\delta_3$ in the $3\nu\text{SM}$ is almost equal to $-\delta_{CP}$. The $ P^{3\nu\text{\tiny{SM}}}_{\odot}$ term is $\mathcal{O}(\Delta^2_{21})$ and small for most experiments we are interested in. The  $P^{\text{\tiny{INT.-$\delta_3$}}}_{\odot-\text{\tiny{ATM.}}}$ term is superficially sensitive to the mass hierarchy, since under $+|\Dmq_{32}|\rightarrow-|\Dmq_{32}|$ it picks up a negative sign and the argument $(|\Dmq_{32}|-\delta_3)\rightarrow (|\Dmq_{32}|+\delta_3)$. Nevertheless, it should be noted that there is no actual sensitivity in the vacuum case, since a rescaling of the CP phase, $\delta_3\rightarrow \pi-\delta_3$, would undo the above transformation~\cite{Minakata:2001qm}.
\par
For later comparison to ``$\,3+2$", in Fig.\,\ref{3neuSMfigs} we make some illustrative plots in the $3\nu\text{SM}$ case, assuming T2K baseline ($295\,\text{Km}$). Note that  in the  $3\nu\text{SM}$ case 
\be
8 |U_{\mu 3}|^2 |U_{e3}|^2=2\sin^2\theta_{23}\sin^22\theta_{13}\;.
\ee
Plotting this combined quantity in Fig.\,\ref{3neuSMfigs} (top left) and later allows us to readily consider a non-maximal atmospheric sector in $|U_{e3}|$ extraction. The CP phase $\delta_3$ now is almost equal to $-\delta_{CP}$, for small $ |U_{e3}|$. 
Note from Fig.\,\ref{3neuSMfigs} (top right) that, for a fixed value of the CP phase, the theoretical differences between NH and IH can be more pronounced as we move away from $\Delta_{32}\sim\pi/2$. This of course does not imply any actual sensitivity to the mass hierarchy in these measurements, due to the invariance under $\delta_3\rightarrow \pi-\delta_3$ and  $+|\Dmq_{32}|\rightarrow-|\Dmq_{32}|$ mentioned earlier~\cite{Minakata:2001qm}.

The bi-probability plot, bottom figure in Fig.\,\ref{3neuSMfigs}, shows the probability orbits in the $(P_{\mu e},\, \overline{P}_{\bar{\mu}\bar{e}})$ plane. The orbits are traced as we vary $\delta_3$, whose values may be read off from the color wheel at the origin, and the size of the ellipses are determined by the magnitude of $|U_{e3}|$. Since $\Delta_{32}\sim\pi/2$ the $\cos\delta_3$ contribution in the interference term is small and the ellipses get squeezed as $\Delta_{32}\rightarrow \pi/2$, tending towards a line~\cite{Barger:2001yr}. Due to this there is no $(\delta_3,\theta_{13})$ degeneracy in the plotted orbits. In this case, if matter matter effects are small (at T2K, for instance, they are small to good approximation), CP-violation can in principle be measured directly by comparing $P_{\mu e}$ and $\overline{P}_{\bar{\mu}\bar{e}}$. 
\begin{figure}
\includegraphics[width=7.5cm,angle=0]{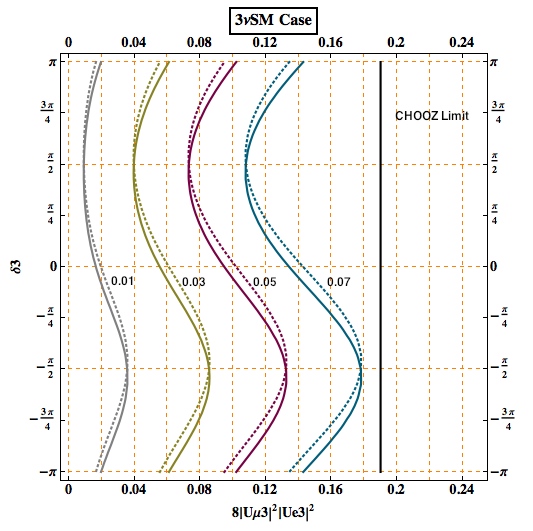}
~~~~~~~~~~~\includegraphics[width=7.5cm,angle=0]{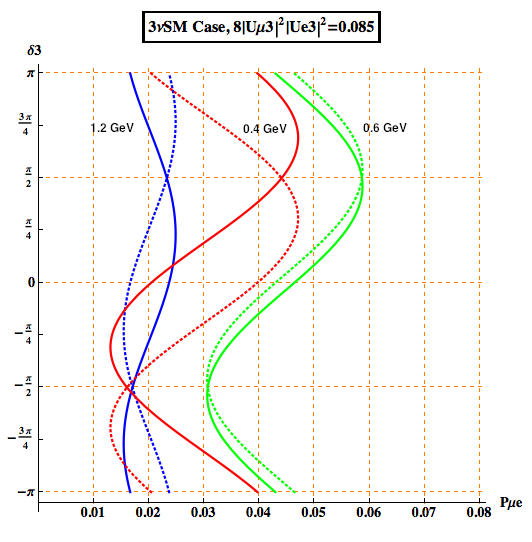}
\includegraphics[width=7.5cm,angle=0]{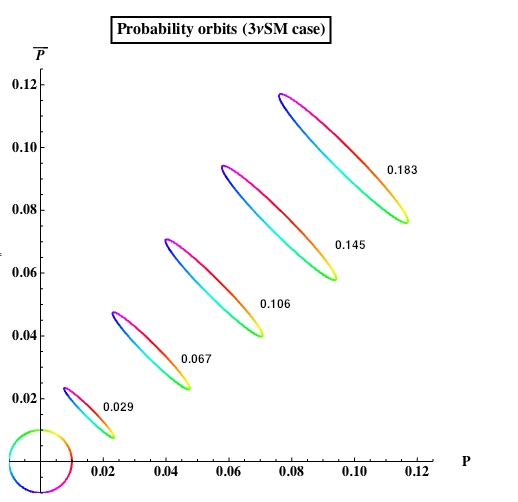}
\caption{Plots in the $3\nu\text{SM}$ case, assuming T2K baseline, showing extracted  $8 |U_{\mu 3}|^2 |U_{e3}|^2$ for $E=0.6\,\text{GeV}$ (top-left plot) and conversion probabilities in the energy bins $\{0.4\,\text{GeV},\,0.6\,\text{GeV},\,1.2\,\text{GeV}\}$ for fixed $8 |U_{\mu 3}|^2 |U_{e3}|^2$ (top-right plot). The thick-lines denote NH and the dotted lines denote IH. The top-left plot is for four fixed probabilities - $0.07,\,0.05,\,0.03\,\text{and}\,0.01$. For comparison, the CHOOZ limit at $0.19$ ($90\%$ C.L.) is shown. To avoid clutter, we have not shown in the plots the best-fit values from the other experiments - $0.041^{+0.047}_{-0.031}$ (MINOS-NH), $0.079^{+0.071}_{-0.053}$ (MINOS-IH), $0.11^{+0.1}_{-0.06}$ (T2K-NH), $0.14^{+0.11}_{-0.08}$ (T2K-IH) and $0.085\pm0.051$ (Double-CHOOZ preliminary)~\cite{{MINOS_theta13}, {Abe:2011sj}, {new-DCHOOZ}}. The bi-probability plot ($\overline{P}_{\bar{\mu}\bar{e}}$ vs. $P_{\mu e}$) for five different values of $4 |U_{e3}|^2 (1-|U_{e3}|^2)$, corresponding to Table \ref{ue3dissvals}, is also plotted for the NH case. The color coding on them denotes the values of $\delta_{3}$ and the values may be ascertained from the color wheel displayed at the origin. Apart from $|U_{e3}|$, for simplicity, wherever applicable all other matrix elements have been assumed to be close to their tribimaximal values.}
\label{3neuSMfigs}
\end{figure}
\par
The first three terms in Eq.\,(\ref{mueconversion}) correspond to these terms above - atmospheric, solar and an interference term between them. The next two terms in Eq.\,(\ref{mueconversion}) are \emph{contributions solely from sterile neutrinos and modulate the $P^{3\nu\text{\tiny{SM}}}_{\text{\tiny{ATM.}}}$ and $ P^{3\nu\text{\tiny{SM}}}_{\odot}$ with relative phase shifts of $\delta_1$ and $\delta_2$ respectively}. Let us label these terms $\Delta P^{\beta''-\delta_1}_{\text{\tiny{ATM.}}}$ and  $\Delta P^{\beta''-\delta_2}_{\odot}$,
\bea
\Delta P^{\beta''-\delta_1}_{\text{\tiny{ATM.}}}&\equiv&~ 4 |U_{\mu 3}| |U_{e3}| |\beta''| \sin\Delta_{31}\sin (\Delta_{31}-\delta_1)\nn\;,\\
\Delta P^{\beta''-\delta_2}_{\odot}&\equiv&~4 |U_{\mu 2}| |U_{e2}| |\beta''| \sin\Delta_{21}\sin (\Delta_{21}-\delta_2)\;.
\eea

The last three terms are energy independent residues obtained after averaging out terms involving large sterile neutrino mass-squared differences. Let us call these residue terms collectively as $\rho^{\text{res.}}$,
\be
\rho^{\text{res.}}\equiv~2\big(|U_{\mu 4}|^2  |U_{e 4}|^2+ |U_{\mu 5}|^2  |U_{e5}|^2+|U_{\mu 4}|  |U^*_{e4}|    |U^*_{\mu 5}|  |U_{e5}| \cos\delta\big)\;.
\label{residue}
\ee

Though higher in order, they become important for small values of $|U_{e3}|$ to give a positive-definite conversion probability. It is also worth emphasizing that both $P^{\text{\tiny{INT.}}-\delta_3}_{\odot-\text{\tiny{ATM.}}}$ and $\Delta P^{\beta''-\delta_1}_{\text{\tiny{ATM.}}}$ are $\propto |U_{e3}|$. This observation will become relevant later when we try to understand cancellations among them.
\par
In Fig.\,\ref{d1d2} we show contour plots of the conversion probability $P_{\mu e}$, for two fixed values of $|U_{e3}|$. T2K base length and characteristic neutrino energy have again been assumed. For the T2K ND distances ($280\,\text{m}$) and $E_{\nu}=0.6\,\text{GeV}$, the effect of sterile neutrinos on the ND fluxes is relatively minimal, but still leads to noticeable spectral distortion. We have included this effect in the analysis and the plots. Focusing primarily on $E_{\nu}\sim0.6\,\text{GeV}$ may be justified by the fact that the J-PARC $\nu_\mu$ beam has a very narrow side band~\cite{Abe:2011sj} and in addition, under ideal conditions, most of the statistical power in $\sin^22\theta_{13}$ extraction may be expected to come from the region of the first oscillation maximum ($\Delta_{32}\sim\pi/2$), tuned at $0.6\,\text{GeV}$. We will look at the effects of varying $E$ later for comparison.
\par
Couple of things may be noted immediately from these contour plots in the $(\delta_1,\delta_2)$ plane, at $\Delta_{32}\sim\pi/2$. The largest conversion to $\nu_e$  takes place in the vicinity of $(0,-\pi/2)$ in both cases. In fact this is found to be true for all intervening values of $|U_{e3}|$ as well. In this region both  $P^{\text{\tiny{INT.}}-\delta_3}_{\odot-\text{\tiny{ATM.}}}$ and $\Delta P^{\beta''-\delta_1}_{\text{\tiny{ATM.}}}$ have the same sign and constructively interfere with  $P^{3\nu\text{\tiny{SM}}}_{\text{\tiny{ATM.}}}$. Heuristically, let us denote this situation as
\be
(0,-\pi/2):P_{\mu e}^{\Delta_{32}\sim\pi/2}\sim P^{3\nu\text{\tiny{SM}}}_{\text{\tiny{ATM.}}}\oplus\left[ P^{\text{\tiny{INT.}}-\delta_3}_{\odot-\text{\tiny{ATM.}}}\oplus\Delta P^{\beta''-\delta_1}_{\text{\tiny{ATM.}}}\right]\;,
\ee
where $\oplus$ denotes constructive interference and $\ominus$ denotes destructive interference. As we decrease $|U_{e3}|$ the conversion probability decreases as expected but the maximal conversion region is relatively unchanged. 
\begin{figure}
\begin{center}
\includegraphics[width=8.5cm,angle=0]{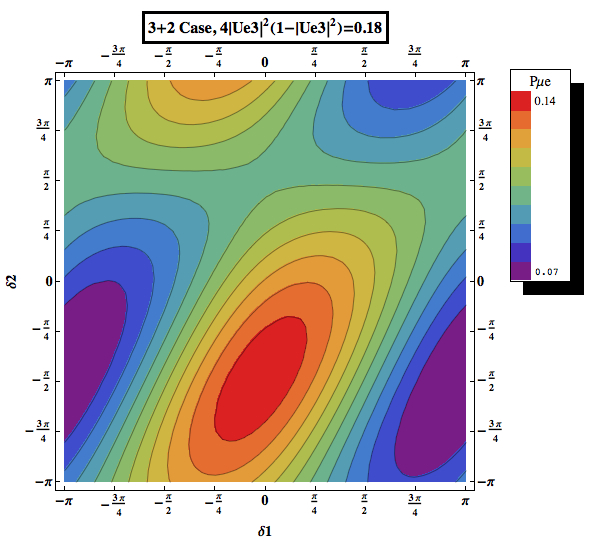}
\includegraphics[width=8.5cm,angle=0]{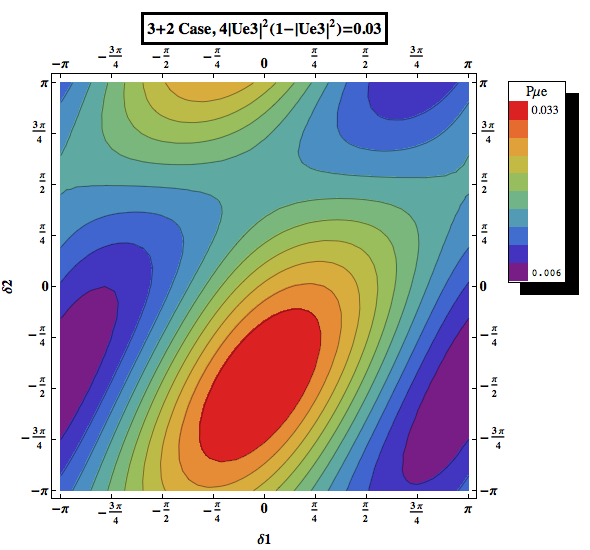}
\end{center}
\caption{Contour plots of $\nu_{\mu}\rightarrow\nu_e$ conversion probability, with T2K parameters, as a function of two independent phases $\delta_1$ and $\delta_2$ near ${\Delta_{32}\sim\pi/2}$ for the NH case. The T2K ND effects have been included. The color-coding denotes the magnitude of the conversion probability in each case. The quantity $ 4 |U_{\mu 3}|^2 (1- |U_{\mu 3}|^2)$ in the $3\nu\text{SM}$ case would have corresponded exactly to $\sin^2 2\theta_{13}$. In the ``$\,3+2$" case, as we had commented earlier, it deviates from $\sin^2 2 \theta_{13}$ by terms of $\mathcal{O}(\theta^2_s)$. In terms of $|U_{e3}|$ the above two plots correspond to $0.22$ (left) and $0.085$ (right), capturing the $|U_{e3}|$ range in Table \ref{ue3dissvals}. In this case, besides $|U_{e3}|$ and the global-fit values of Table \ref{globalfit}, again all other matrix elements have been assumed to be close to their tribimaximal values.}
\label{d1d2}
\end{figure}
\par
Similarly, the lowest conversion probabilities occur in the vicinity of $(\pm\pi,-\pi/2)$, where $P^{\text{\tiny{INT.}}-\delta_3}_{\odot-\text{\tiny{ATM.}}}$ and $\Delta P^{\beta''-\delta_1}_{\text{\tiny{ATM.}}}$ have the same sign but now destructively interfere with  $P^{3\nu\text{\tiny{SM}}}_{\text{\tiny{ATM.}}}$,
\be
(\pm\pi,-\pi/2):P_{\mu e}^{\Delta_{32}\sim\pi/2} \sim ~P^{3\nu\text{\tiny{SM}}}_{\text{\tiny{ATM.}}} \ominus \left[ P^{\text{\tiny{INT.}}-\delta_3}_{\odot-\text{\tiny{ATM.}}}\oplus\Delta P^{\beta''-\delta_1}_{\text{\tiny{ATM.}}}\right]\;.
\ee
\par
 Probably even more interesting is the observation of a thin band near $\delta_2=\pi/2$ for which the conversion probability is almost constant over the full range of $\delta_1$. In this band there is an almost perfect cancellation between the $P^{\text{\tiny{INT.}}-\delta_3}_{\odot-\text{\tiny{ATM.}}}$ and $\Delta P^{\beta''-\delta_1}_{\text{\tiny{ATM.}}}$ terms,
\be
(\forall  \delta_1,+\pi/2):P_{\mu e}^{\Delta_{32}\sim\pi/2} \sim ~P^{3\nu\text{\tiny{SM}}}_{\text{\tiny{ATM.}}}\oplus\left[ P^{\text{\tiny{INT.}}-\delta_3}_{\odot-\text{\tiny{ATM.}}}\ominus\Delta P^{\beta''-\delta_1}_{\text{\tiny{ATM.}}}\right]\; .
\label{d2ppiby2}
\ee
 The conversion probability in this case should almost be identical to the conversion probability in the $3\nu\text{SM}$ case, when the interference term $P^{\text{\tiny{INT.}}-\delta_3}_{\odot-\text{\tiny{ATM.}}}$ in the $3\nu\text{SM}$ limit almost completely vanishes (i.e. ($\Delta_{32}-\delta_3)\,\simeq \pm\pi/2$ in Eq. (\ref{3nsmconv})). In the $3\nu\text{SM}$ case, near $\Delta_{32}\sim\pi/2$, this happens at $\delta_3=0$ and $\delta_3=\pm\pi$. The conversion probabilities are indeed found to match as expected upon comparison. Again, for the global-fit and close-to-tribimaximal values we are working with, \emph{this conclusion is seen to be true, independent of $U_{e3}$, since both $P^{\text{\tiny{INT.}}-\delta_3}_{\odot-\text{\tiny{ATM.}}}$ and $\Delta P^{\beta''-\delta_1}_{\text{\tiny{ATM.}}}$ are $\propto |U_{e3}|$}.
\par
When $\delta_2=0$, it is seen that the terms $P^{\text{\tiny{INT.}}-\delta_3}_{\odot-\text{\tiny{ATM.}}}$ and $\Delta P^{\beta''-\delta_1}_{\text{\tiny{ATM.}}}$ are out of phase with each other by $\pi/2$ in the phase $\delta_1$ (or now equivalently $\delta_3$). Due to this, the conversion probability attains its maximum and minimum values at $\delta_3=\pi/4$ and $\delta_3=-3\pi/4$ respectively when $\Delta_{32}\sim\pi/2$.
\begin{figure}
\begin{center}
\includegraphics[width=7.0cm,angle=0]{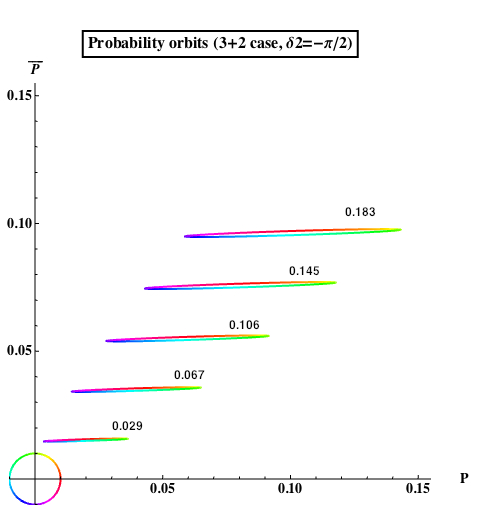}
~~~~~~~~~~~~\includegraphics[width=7.0cm,angle=0]{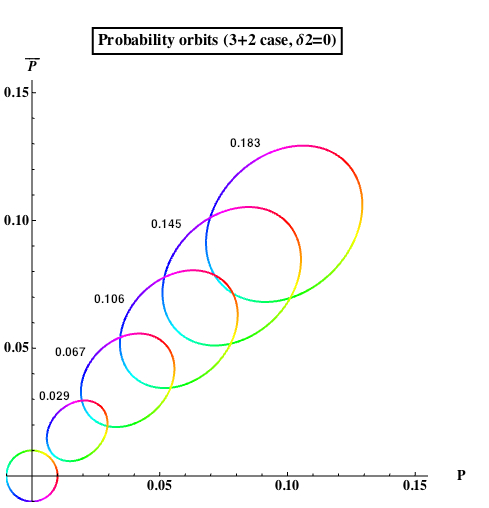}
\includegraphics[width=7.0cm,angle=0]{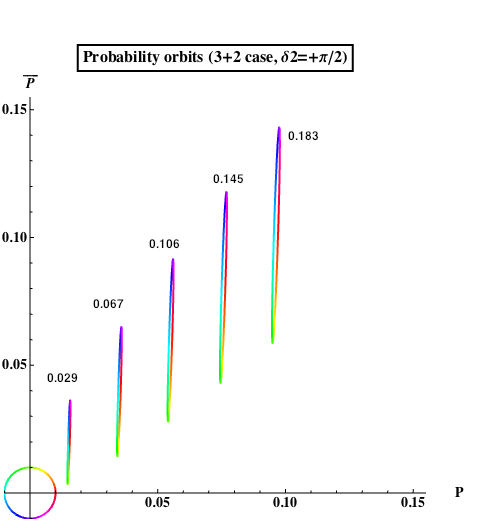}
\end{center}
\caption{Probability orbits in the ``$\,3+2$" neutrino scenario for the NH case. The respective values of $4 |U_{e3}|^2 (1-|U_{e3}|^2)$ are labelled on the orbits. As before, apart from $|U_{e3}|$ and the global-fits, all other matrix elements have been assumed to be close to their tribimaximal values. It is observed that in contrast to the $3\nu\text{SM}$ case, the $(\delta_3,\theta_{13})$ degeneracy may be re-introduced depending on the value of the phase $\delta_2$. Also observe that $\delta_2=-\pi/2$ and $\delta_2=+\pi/2$ are orthogonal choices for which $\overline{P}_{\bar{\mu}\bar{e}}$ and $P_{\mu e}$ remain almost constant respectively, as $\delta_3$ traces the orbit. These may be compared to the $3\nu\text{SM}$ probability orbits in Fig.\,\ref{3neuSMfigs}. The value of $\delta_3$ at any point in the orbit may again be deduced from the color wheel at the origin.}
\label{SterileProbOrbits}
\end{figure}
\par
In Fig.\,\ref{SterileProbOrbits} we look at the probability orbits in the $(P_{\mu e},\, \overline{P}_{\bar{\mu}\bar{e}})$ plane for $\Delta_{32}\sim\pi/2$. We see a rich behavior in the orbits depending on the value of $\delta_2$. These may again be understood in terms of interference between $P^{\text{\tiny{INT.}}-\delta_3}_{\odot-\text{\tiny{ATM.}}}$ ($\overline{P}^{\text{\tiny{INT.}}-\delta_3}_{\odot-\text{\tiny{ATM.}}}$) and $\Delta P^{\beta''-\delta_1}_{\text{\tiny{ATM.}}}$ ($\Delta \overline{P}^{\beta''-\delta_1}_{\text{\tiny{ATM.}}}$), in $P_{\mu e}$ ($\overline{P}_{\bar{\mu}\bar{e}}$). Since $P(\bar{\nu}_\alpha \rightarrow \bar{\nu}_\beta;U)=P(\nu_\alpha \rightarrow \nu_\beta;U^*)$, all the phases change sign as we go from neutrinos to anti-neutrinos. Specifically as seen from Eq.\,(\ref{d2ppiby2}), for $\delta_2=-\pi/2$, the $\overline{P}^{\text{\tiny{INT.}}-\delta_3}_{\odot-\text{\tiny{ATM.}}}$ and $\Delta \overline{P}^{\beta''-\delta_1}_{\text{\tiny{ATM.}}}$ terms cancel each other for $\overline{P}_{\bar{\mu}\bar{e}}$ leaving a residue almost independent of $\delta_3$. This may be symbolically expressed as
\bea
(\forall  \delta_1,-\pi/2):\overline{P}_{\bar{\mu} \bar{e}}^{\Delta_{32}\sim\pi/2} &\sim& ~P^{3\nu\text{\tiny{SM}}}_{\text{\tiny{ATM.}}}\oplus\left[ \overline{P}^{\text{\tiny{INT.}}-\delta_3}_{\odot-\text{\tiny{ATM.}}}\ominus\Delta \overline{P}^{\beta''-\delta_1}_{\text{\tiny{ATM.}}}\right]\; .
\eea
It is also interesting to note that for $\delta_2=0$ the $(\delta_3,\theta_{13})$ degeneracy is reintroduced even when $\Delta_{32}\sim\pi/2$. This is clearly seen from the fact that the elliptic orbits for adjacent $|U_{e3}|$ values intersect. At the points of intersection both $\overline{P}_{\bar{\mu}\bar{e}}$ and $P_{\mu e }$ have the same magnitudes for different values of $\delta_3$ and $|U_{e3}|$. This degeneracy \textit{can now mix CP-conserving and CP-violating solutions}. An example of this may be observed in the $\delta_2=0$ case of Fig.\,\ref{SterileProbOrbits} where the orbit labelled by $0.183$  intersects the orbit labelled by $0.106$. The former solution is CP-violating ($\delta_3\neq0$) whereas the latter is CP-conserving ($\delta_3\simeq0$). Note that in the bi-probability discussions, to first approximation, the effects of the base length ($L$) and neutrino energy ($E$) only appear through $\Delta_{32}\sim\pi/2$. Due to this the general features of the probability orbits should be more widely valid, as long as we are in the vicinity of a conversion maximum. 
\begin{figure}
\begin{center}
\includegraphics[width=7.5cm,angle=0]{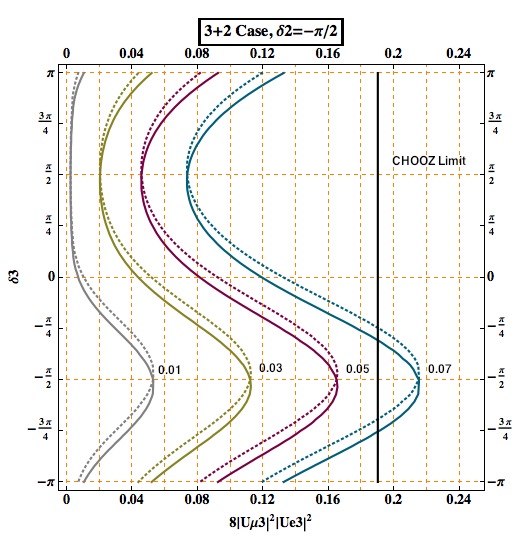}
~~~~~~~~~~~~\includegraphics[width=7.5cm,angle=0]{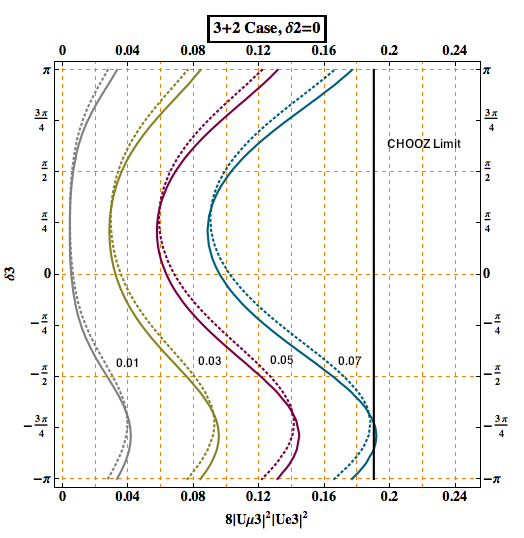}
\includegraphics[width=7.5cm,angle=0]{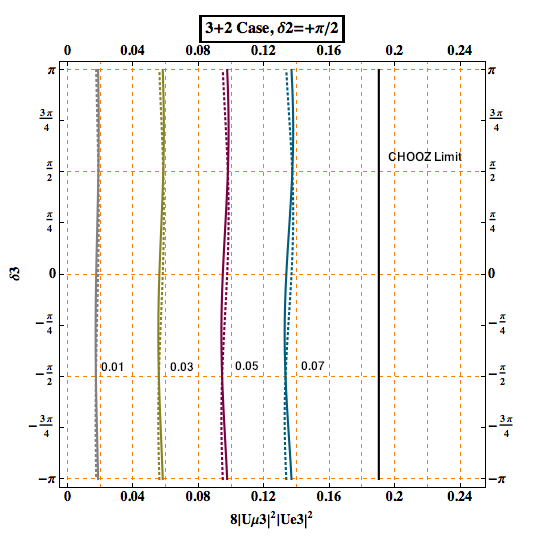}
\end{center}
\caption{Extracted values of $8 |U_{\mu 3}|^2 |U_{e3}|^2$ in the ``$\,3+2$" case, with T2K parameters ($L=295\,\text{Km}$, $E=0.6\,\text{GeV}$), assuming fixed conversion probabilities - $0.07,\,0.05,\,0.03,$ and $0.01$. The thick-lines are for NH  and the dotted-lines are for IH. We re-emphasize that the quantity $8 |U_{\mu 3}|^2 |U_{e3}|^2$ plotted would exactly correspond to $2\sin^2\theta_{23}\sin^22\theta_{13}$  in the  $3\nu\text{SM}$. The other matrix elements have been chosen as in the previous cases. For comparison, the preliminary best-fit value from Double-CHOOZ is at $0.085\pm0.051$~\cite{{new-DCHOOZ}}.}
\label{Ue3extsterile}
\end{figure}
\par
Fig.\,\ref{Ue3extsterile} shows the extracted values of $8 |U_{\mu 3}|^2 |U_{e3}|^2$ for fixed conversion probabilities, again assuming that most of the statistical significance is coming from the region near $\Delta_{32}\sim\pi/2$. The thick-lines indicate NH and the dotted-lines indicate IH for comparison. 
\par
The different values of $8 |U_{\mu 3}|^2 |U_{e3}|^2$ extracted in Fig.\,\ref{Ue3extsterile} are easily understood by looking at the corresponding conversion probabilities near that particular CP phase region. For a fixed conversion probability, positive (negative) interference terms from $P^{\text{\tiny{INT.}}-\delta_3}_{\odot-\text{\tiny{ATM.}}}$ and $\Delta P^{\beta''-\delta_1}_{\text{\tiny{ATM.}}}$ must be compensated by lower (higher) value of $|U_{e3}|$. Therefore, the variation of the extracted $|U_{e3}|$ magnitude as we vary $\delta_3$ must be anti-correlated with the conversion probability variation. 
\begin{figure}
\begin{center}
\includegraphics[width=7.5cm,angle=0]{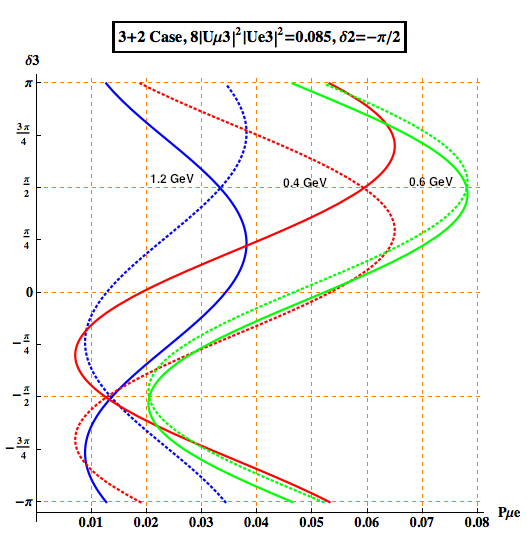}
~~~~~~~~~~~\includegraphics[width=7.65cm,angle=0]{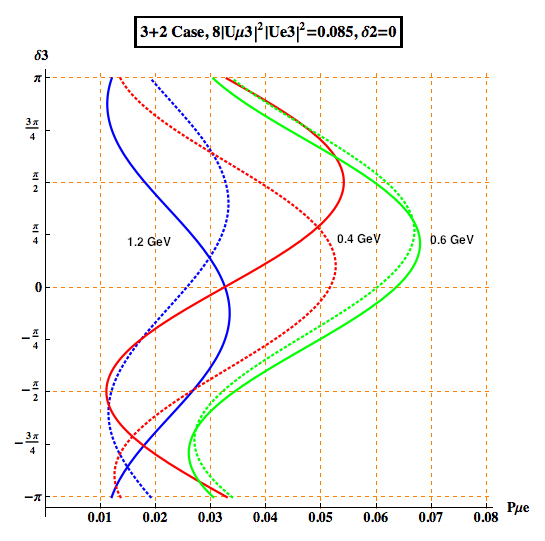}
\includegraphics[width=7.5cm,angle=0]{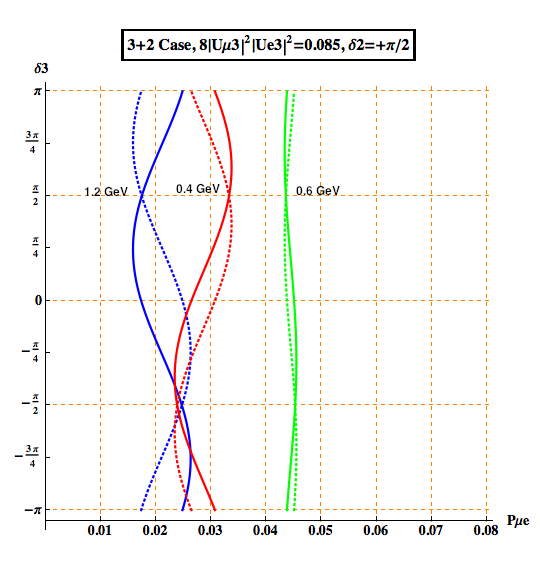}
\end{center}
\caption{Conversion probabilities $P_{\mu e}$ in the ``$\,3+2$" case, for three different neutrino energies - $0.4\,\text{GeV}$ (red), $0.6\,\text{GeV}$ (green) and $1.2\,\text{GeV}$ (blue). The thick-lines are for NH and the dotted-lines are for IH as before. They are plotted for a fixed $8 |U_{\mu3}|^2 |U_{e3}|^2$ of $0.085$. It is clear that there may be significant differences between NH and IH depending on $E$ and the CP phase structure.}
\label{Espread}
\end{figure}
\par
Another point we would like to emphasize is that the actual $|U_{e3}|$ value extracted depends on the assumed value of $|U_{\mu3}|$, which may be extracted, as we saw in Eq.\,(\ref{musurv3p2gen}), from a $\nu_\mu$ disappearance measurement such as MINOS. We saw in the case of LBL, and specifically MINOS, that  the presence of sterile neutrinos cause the  value of extracted $|U_{\mu3}|$ also to shift to lower values by a few percent
\be
 |U_{\mu3}|^{3+2}_{\text{\tiny{extr.}}}~\lesssim~|U_{\mu3}|^{3\nu\text{SM}}_{\text{\tiny{extr.}}} \; .
\ee
Due to these considerations, plotting $8 |U_{\mu 3}|^2 |U_{e3}|^2$ in Fig.\,\ref{Ue3extsterile} allows us to incorporate a non-maximal atmospheric sector and $|U_{\mu 3}|$ readily. 
Let us try to understand some of the theoretical features in the plots.
From Fig.\,\ref{3neuSMfigs} we observe that in the $3\nu\text{SM}$ case the extracted $|U_{e3}|$ is the same for normal and inverted hierarchy when $\delta_3=\pm\pi/2\, \forall \, E$. This is due to the fact that the interference term becomes $\propto\pm\sin^2\Delta_{31}$ for  $\delta_3=\pm\pi/2$. It is also clear that the maximum difference between normal and inverted hierarchies occur at $\delta_3=\{0,\pm \pi\} \,\forall \, E$, as expected from Eq.\,(\ref{3nsmconv}).
It is worth re-emphasizing that these theoretical features again do not imply actual experimental sensitivities to neutrino mass hierarchy, in vacuum, due to the invariance under $\delta_3\leftrightarrow\pi-\delta_3$.
\par
Along similar lines we can understand the features in the ``$\,3+2$" case. The NH and IH cases should give the same extracted $|U_{e3}|$ when the terms which transform under $|\Dmq_{32}|\rightarrow-|\Dmq_{32}|$, specifically $P^{\text{\tiny{INT.}}-\delta_3}_{\odot-\text{\tiny{ATM.}}}$ and $\Delta P^{\beta''-\delta_1}_{\text{\tiny{ATM.}}}$, sum to the same numerical value, apart from the $|U_{e3}|$ factor that is common to both. If a term by term equivalence is demanded (which is a stronger condition than required) between NH and IH, we must have
\bea
\left[\cos (|\Delta_{32}|-\delta_3)+\cos (|\Delta_{32}|+\delta_3)\right]&\rightarrow&0\; ,\nn\\
\left[\sin(|\Delta_{31}|+\delta_1)-\sin (|\Delta_{31}|-\delta_1)\right]&\rightarrow&0\; ,
\eea
These give the solutions in the $(\delta_1,\delta_2)$ space :
\bea
\text{NH} \equiv \text{IH}~\forall \,(L,\,E)~:~(0,\pm\pi/2),\,(\pi,\pm\pi/2),\,(-\pi,\pm\pi/2)\;.~~~~~~
\eea
These regions are clearly visible in the top-left ($\delta_2=-\pi/2$) and bottom ($\delta_2=+\pi/2$) plots of Fig.\,\ref{Ue3extsterile}, with the understanding that $\delta_3=\delta_1-\delta_2$.
\par
Imposing the weaker condition that the net sum of  $P^{\text{\tiny{INT.}}-\delta_3}_{\odot-\text{\tiny{ATM.}}}$ and $\Delta P^{\beta''-\delta_1}_{\text{\tiny{ATM.}}}$ be equivalent in the NH and IH cases leads to the solution
\bea
\text{NH} \equiv \text{IH}\,\forall \,(L,\,E):\frac{\sin\delta_1}{\cos(\delta_1-\delta_2)}\simeq
\frac{2 |U_{\mu 2}| |U^{*}_{e2}| \sin\Delta_{21}\cos|\Delta_{32}|}{|\beta''|\cos|\Delta_{31}|}.
\eea
This is independent of $|U_{e3}|$ and $|U_{\mu 3}|$.   For $E=0.6\,\text{GeV}$ ($\Delta_{32}\sim\pi/2$ for $L_{\text{\tiny{T2K}}}$), $(\delta_1,\delta_2) \sim (\pi/4,0)$ and $(-3\pi/4,0)$ are among the approximate solutions to the above equation. This equivalence between NH and IH can be seen clearly in the top-right plot of Fig.\,\ref{Ue3extsterile} near these regions.
\par
For most of our theoretical discussions till this point, we were focused on the region near $\Delta_{32}\sim\pi/2$ at T2K, and hence at neutrino energies near $600\,\text{MeV}$. As perviously noted, it may be argued that this is not too egregious a choice since the J-PARC $\nu_\mu$ beam has a very narrow side band and in addition, naively, most of the statistical power in $\sin^22\theta_{13}$ extraction must come from the region of the first oscillation maximum, tuned at $0.6\,\text{GeV}$. Nevertheless it is important to explore the variations with $E$, especially considering that the T2K $\nu_e$ appearance measurement observed 4 events outside the $0.6\,\text{GeV}$ bin.
\par
Fig.\,\ref{Espread} shows the variation in the conversion probability $P_{\mu e}$ as we vary $E$ for NH and IH. They are plotted for a fixed $8 |U_{\mu3}|^2 |U_{e3}|^2$ of $0.085$. We focus on the $400\,\text{MeV}$,  $600\,\text{MeV}$ and $1200\,\text{MeV}$ energy bins. These bins correspond to $\Delta_{32}\sim3\pi/4$, $\Delta_{32}\sim\pi/2$ and $\Delta_{32}\sim\pi/4$ for the T2K baseline. As is clear, the variations may be substantial between NH and IH as we move away from the oscillation maximum depending on $(\delta_1,\delta_2)$. We could now pose the question - for a fixed conversion probability near $\Delta_{32}\sim\pi/2$ what is the smallest $|U_{e3}|$ it may be associated with for any CP phase structure in the $3\nu\text{SM}$ and ``$\,3+2$" cases ?
\par
In Fig.\,\ref{Ue3min} we show the minimum attainable $|U_{e3}|$ values, in terms of $4 |U_{e3}|^2 (1-|U_{e3}|^2)$, in the $3\nu\text{SM}$ and ``$\,3+2$" scenarios. The range of conversion probabilities in Fig.\,\ref{Ue3min} correspond to those in Fig.\,\ref{Ue3extsterile}. Note that \emph{the smallest possible $|U_{e3}|$ in the ``$\,3+2$" case is always significantly smaller than that possible in the $3\nu\text{SM}$,}
\be
|U_{e3}|^{3\nu\text{SM}}_{\text{\tiny{smallest}}}~>~ |U_{e3}|^{3+2}_{\text{\tiny{smallest}}}\;.
\ee 
 \emph{Also observe that, though reduced, the smallest $|U_{e3}|$ values in ``$\,3+2$" are still different from zero, albeit extremely tiny for small conversion probabilities}. This is because, $P_{\odot}$ and $\Delta P^{\beta''-\delta_2}_{\odot}$ by themselves are not sufficient to fulfill a given conversion probability in the range shown. 
\par
In the 3-neutrino scenario, from the T2K lower-bound curves~\cite{Abe:2011sj} (which give the $90 \%$ C.L. lower bound on $\sin^22\theta_{13}$ as $0.03-0.04$ for $\delta_{\text{\tiny{CP}}}=0$), the smallest possible $\sin^22\theta_{13}$ is about $0.02-0.025$, at $\delta_{\text{\tiny{CP}}}\simeq-\pi/2$. If we assume that the lower-bound curves correspond approximately to constant conversion probabilities, then using those approximate $P_{\mu e}$ values in the ``$\,3+2$" scenario we may estimate,  from Fig.\,\ref{Ue3min}, a lower limit for $4 |U_{e3}|^2 (1-|U_{e3}|^2)$. This gives for the ``$\,3+2$" case,
\be
4 |U_{e3}|^2 (1-|U_{e3}|^2)~\gtrsim ~~0.008-0.01; \;\;\;\; (|U_{e3}| ~\gtrsim ~~0.04-0.05) \;\;\; {\rm at \; 90 \% \; \text{C.L.}}~~~\;.
\ee
 Based on the above discussion, we note that in the ``$\,3+2$" case, the results from T2K imply a $90 \%$ C.L. lower bound that is still within the reach (sensitivity) of future reactor neutrino experiments like Daya Bay~\cite{Wang:2011tp}, and consistent with the $1\sigma$ range of $\sin^22\theta_{13}$ recently reported by the Double-CHOOZ experiment. 


\begin{figure}
\includegraphics[width=11.25cm,angle=0]{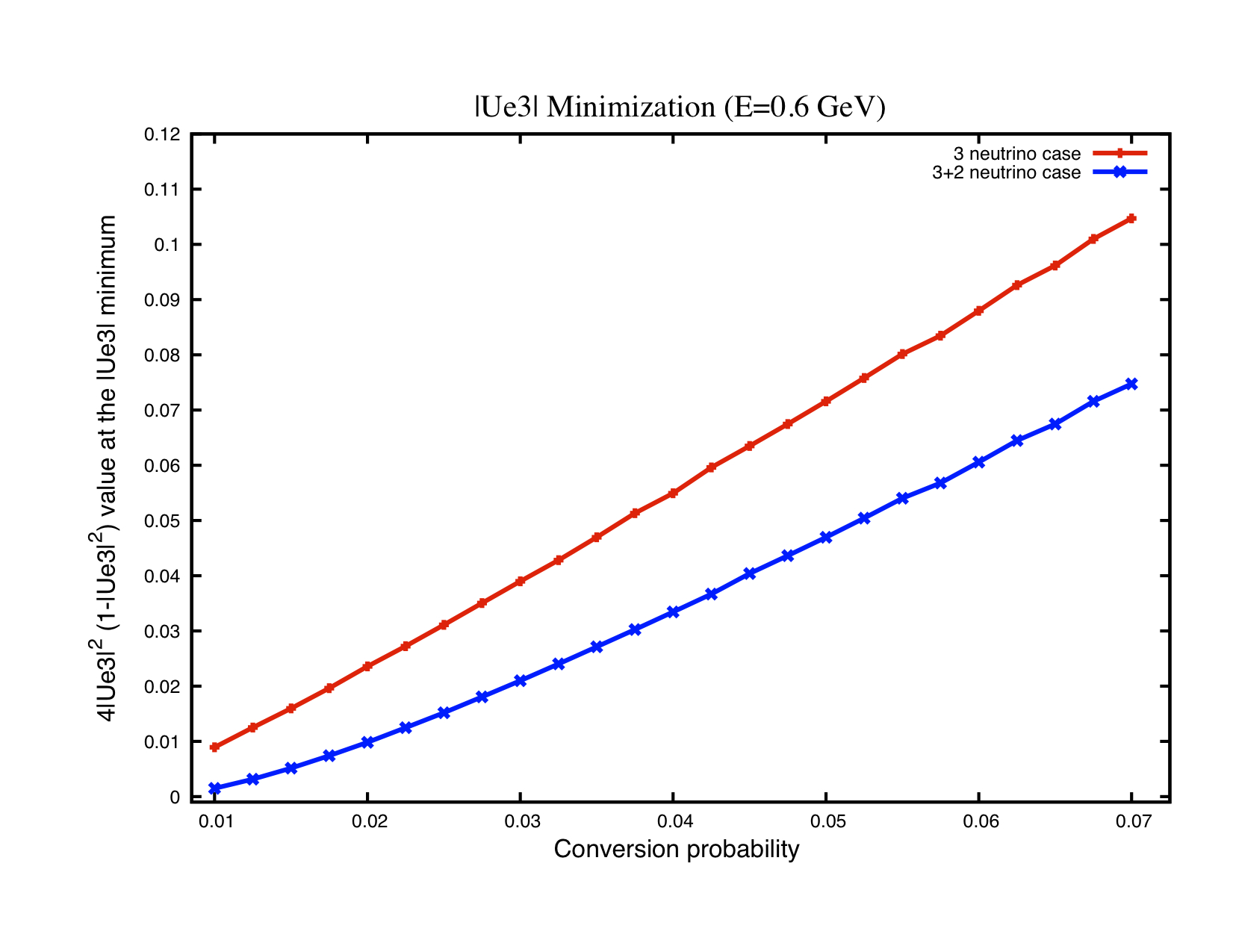}
\caption{Minimum possible $|U_{e3}|$  in the $3\nu\text{SM}$ and ``$\,3+2$" cases (NH and IH) expressed in terms of $4 |U_{e3}|^2 (1-|U_{e3}|^2)$. In the plot, $E$ is fixed at $0.6\,\text{GeV}$, the implicit assumption being that most of the statistical power in $|U_{e3}|$ extraction may come from the vicinity of $E\sim 0.6\,\text{GeV}$ (equivalently $\Delta_{32}\sim\pi/2$ for T2K baseline) which is the oscillation maximum. It is clear that the lower-bounds on the extracted  $|U_{e3}|$ are generally much smaller with a ``$\,3+2$" assumption, but still non-zero. The best-fit values in the $3\nu\text{SM}$ for comparison are  - $0.041^{+0.047}_{-0.031}$ (MINOS-NH), $0.079^{+0.071}_{-0.053}$ (MINOS-IH), $0.11^{+0.1}_{-0.06}$ (T2K-NH), $0.14^{+0.11}_{-0.08}$ (T2K-IH) and $0.085\pm0.051$ (Double-CHOOZ preliminary)~\cite{{MINOS_theta13}, {Abe:2011sj}, {new-DCHOOZ}}.}
\label{Ue3min}
\end{figure}

\subsubsection {\bf Matter effects}

\par
All the above effects, due to the presence of sterile neutrinos, may be further modified by matter effects, depending on the base length and $E$.
In the $3\nu\text{SM}$ case these matter effects may be quantified as~\cite{Akhmedov:2004ny}
\bea
\mathcal{P}^{3\nu\text{SM}}_{ee}&=&1  - 4 s_{13}^2  \frac{\sin^2 (A_\text{\tiny{M}}-1)\Delta_{31}}{(A_\text{\tiny{M}}-1)^2}- \epsilon^2  \sin^2 2\theta_{12} \frac{\sin^2
  A_\text{\tiny{M}}\Delta_{31}}{A_\text{\tiny{M}}^2} \,,\nn \\
\mathcal{P}^{3\nu\text{SM}}_{e\mu}&=&4 s_{13}^2 s_{23}^2 \frac{\sin^2 (A_\text{\tiny{M}}-1)\Delta_{31}}{(A_\text{\tiny{M}}-1)^2}+2 \epsilon \,s_{13}\,\sin 2\theta_{12}\sin2\theta_{23} \cos(\Delta_{31} - \delta_{\rm CP})\frac{\sin A_\text{\tiny{M}}\Delta_{31}}{A_\text{\tiny{M}}} \frac{\sin (A_\text{\tiny{M}}-1)\Delta_{31}}{A_\text{\tiny{M}}-1} \nn\\
&+& \epsilon^2 \sin^2 2\theta_{12} c_{23}^2 \frac{\sin^2
 A_\text{\tiny{M}}\Delta_{31}}{A_\text{\tiny{M}}^2} \;,
  \label{mattereffs}
\eea
where
\be
\epsilon=\frac{\Delta m^2_{21}}{\Delta m^2_{31}}~~,~~A_\text{\tiny{M}}=\frac{2E V_\text{\tiny{M}}}{\Dmq_{31}}\;.
\ee
Eq.\,(\ref{mattereffs}) is written to second order in $\epsilon$ and $\sin\theta_{13}$, assuming a constant matter-density potential~\cite{Akhmedov:2004ny}
\be
V_\text{\tiny{M}}\simeq7.56\times 10^{-14}\left(\frac{\rho_{\text{\tiny{crust}}}}{\text{g}/\text{cm}^3}\right)Y_{e}~~~\text{eV}\;,
\ee
where $\rho_{\text{\tiny{crust}}}$ is the crust matter density and $Y_e$ is the number of electrons per nucleon. For earth matter $Y_e\simeq0.5$ to very good approximation.
\par
For a constant earth-crust density $\rho_{\text{\tiny{crust}}}\simeq\,3 \,\text{g}/\text{cm}^3$, we can estimate using the characteristic experimental parameters that
\bea
&A^\text{\tiny{D-CHOOZ}}_\text{\tiny{M}}&\simeq~0.0003\nn\; ,\\
&A^\text{\tiny{T2K}}_\text{\tiny{M}}~~~&\simeq~0.06\nn\; ,\\
&A^\text{\tiny{MINOS}}_\text{\tiny{M}}&\simeq~0.3\;.
\eea
Using the above values and Eq.\,(\ref{mattereffs}) we can make estimates to convince ourselves that for Double-CHOOZ the matter effects are almost completely irrelevant and the extracted $|U_{e3}|$ is hardly affected. For T2K, it is seen that the matter effects are still relatively minimal (near $E\sim0.6\,\text{GeV}$) but induce at most a few percent change in the extracted  $|U_{e3}|$ relative to the vacuum assumption. In MINOS the matter effects can become more significant and may induce larger modifications of the extracted $|U_{e3}|$ somewhat obscuring any possible additional effects due to sterile neutrinos. 
\par
In both cases above, for a fixed conversion probability, the effect of matter interactions is to \emph{decrease} (\emph{increase}) the extracted $|U_{e3}|$ for NH (IH). This is probably most easily understood in a 2-neutrino limit by noting that the effect of the matter potential is to increase (decrease) the effective $\sin^22\theta_{\text{\tiny{M}}}$ coefficient for NH (IH). Crudely, to lowest order,  the above conclusion should still hold approximately in the ``$\,3+2$" neutrino case. Also note that in contrast to matter effects, the direction in which $|U_{e3}|$ was modified due to sterile neutrinos depended intricately on the $(\delta_1,\delta_2)$ CP phase structure. A comprehensive analysis of $\theta_{13}$ extraction including matter-effects at MBL/LBL, in the presence of two sterile neutrinos, is beyond the scope of the present work~(see \cite{Karagiorgi:2011ut} and references therein in this context, for a ``$\,3+1$" SBL fit incorporating matter effects).

\begin{subsubsection} {\bf Comparison to the ``$\,3+1$" case.}
It is interesting to point out in the MBL/LBL limit that for $P_{\mu e}$, if one were to use the best-fit values in ``$\,3+1$" for the matrix elements, as in \cite{Giunti:2011ht} say, the numerical values of $\beta''$ and the energy independent residue term $\rho^{\text{res.}}$ defined in Eq.\,(\ref{residue}) comes out to be numerically almost the same. In the ``$\,3+2$"  case we have
\bea
|\beta^{''}|_{\text{``}3+2\text{"}}&=&0.0351\nn\;,\\
\rho^{\text{res.}}_{\text{``}3+2\text{"}}&=&0.0021\;
\eea
and in the ``$\,3+1$"  case we get for the equivalent values,
\bea
|\beta^{''}|_{\text{``}3+1\text{"}}&=&0.0354\nn\;,\\
\rho^{\text{res.}}_{\text{``}3+1\text{"}}&=&0.0025\; .
\eea
\par
These quantities along with the 2 independent phases are the only relevant quantities in the LBL/MBL limit that depend on the presence of sterile neutrinos. The larger $\Delta m^2_{41}$ one obtains from SBL fits in ``$\,3+1$", relative to ``$\,3+2$" $\Delta m^2$ values, is irrelevant for MBL/LBL since the terms containing it get averaged at the FD anyway. 
\end{subsubsection}
\end{subsection}
\end{section}

\begin{section}{Summary and Conclusions}
In the present study, we revisited some of the recent neutrino observations in the context of sterile neutrinos and the global fits from SBL experiments, to understand their impact on current and upcoming MBL/LBL measurements.
\par
We noted that in general, for LBL experiments, the existence of sterile neutrinos lead to a distinct parametrization of the oscillation survival probabilities in terms of a normalization factor and a modified coefficient of the energy dependent term. We analyzed the MINOS neutrino and anti-neutrino disappearance data~\cite{{Adamson:2011ig},{Adamson:2011fa},{Adamson:2011ch}} from this perspective. Though the parametrization does lead to a marginal improvement in fit, it was found that the current MINOS data by itself does not  definitively discriminate the ``$\,3+2$" scenario or the parameter values obtained from SBL fits. It was found that the $|U_{\mu 3}|$ confidence interval shifts to lower values by a few percent when the possible existence of sterile neutrinos are taken into account.
\par
It was also commented that the recent measurements of a possibly non vanishing reactor angle $\theta_{13}$ may be affected by the existence of sterile neutrinos. We pointed out that the existence of sterile neutrinos may induce a modification of this angle (more precisely $|U_{e3}|$) in experiments that look at neutrino conversion probabilities, such as T2K and MINOS, and the perceived value may be shifted significantly from the ``true" value in these cases. We also studied in detail the effects of additional sterile neutrino terms and their interference due to CP phases, in the ``$\,3+2$" conversion probabilities.  The probability orbits in the bi-probability plots also exhibited interesting features distinct from $3\nu\text{SM}$. It was, for instance, observed that the $(\delta_3,\theta_{13})$ degeneracy may be re-introduced depending on the CP phase structure in the ``$\,3+2$" scenario and that there may be orbits where either the $\overline{P}_{\bar{\mu}\bar{e}}$ or $P_{\mu e}$ value remains almost constant with changing $\delta_3$.
\par
It was also reiterated in the study that in the reactor experiments, these modifications due to sterile states are less significant. Due to this, the matrix element $|U_{e3}|$ when determined from survival probabilities under the $3\nu\text{SM}$ assumption, is close to the ``$\,3+2$" value, as compared to when determined from conversion probabilities. Neutrino disappearance experiments include Double-CHOOZ~\cite{Akiri:2011zz} and upcoming experiments such as Daya Bay~\cite{Wang:2011tp} and RENO~\cite{Jeon:2011zz} that will measure $\theta_{13}$ to high precision. In this context we also conclude from our study that the results from T2K imply a $90 \%$ C.L.  lower-bound on $|U_{e3}|$, in the ``$\,3+2$" neutrino case, which is still within the sensitivity of future reactor neutrino experiments like Daya Bay~\cite{Wang:2011tp}, and consistent with the one-$\sigma$ range of $\sin^22\theta_{13}$ recently reported by the Double-CHOOZ experiment. Finally, we argued that the results in the ``$\,3+1$"  scenario, using the recent best-fit values, would be very close to the medium/long baseline results we obtained in the ``$\,3+2$" case. This was attributed to the numerical equivalence of the relevant parameters in both cases.
\par
Our analysis suggests that if the SBL global fits, including the anomalies, are in fact legitimate indications of sterile neutrinos in nature, then there may be interesting effects in MBL/LBL neutrinos experiments. A more comprehensive study in the ``$\,3+2$" scenario including matter effects at MBL/LBL terrestrial neutrino experiments, such as MINOS~\cite{MINOS_theta13} and No$\nu$A~\cite{Davies:2011vd}, is left for future exploration. We also plan to pursue in future, a study of how the ``$\,3+2$" scenario affects survival probabilities for solar neutrinos~\cite{Giunti:2009xz}, using current global fit parameters.
\par
\vspace{1mm}
\textbf{Note added :} After the submission of this article a similar work \cite{Giunti:2011vc} appeared that discusses the recent Double-CHOOZ results considering the energy dependence of the events at  the ND induced by the presence of sterile neutrinos. The final conclusion is similar to ours namely that the value of $\sin^2 2\theta_{13}$ is not significantly modified with respect to the three neutrino case, although due to the sterile neutrino effects the final uncertainties associated with the result are somewhat larger than the ones quoted by the Double-CHOOZ experiment.
\end{section}



\begin{acknowledgments}
We thank E. Blucher, Z. Djurcic, J. Evans, G. O. Gann, M. Goodman, J. Kopp and M. Sanchez for discussions. C.W would like to acknowledge discussions with I. Mocioiu during an early investigation that partly motivated the present work. Work at ANL is supported in part by the U.S. Department of Energy (DOE), Div. of HEP, Contract DE-AC02-06CH11357. B.B and A.T were supported in part by the United States Department of Energy through Grant No. DE-FG02-90ER40560. A.T also acknowledges support from the Sidney Bloomenthal Fellowship during latter stages of this work.
\end{acknowledgments}

\begin{appendix}
\section{Matrix elements}
For completeness, we list some of the relevant matrix elements in terms of the angles and phases in a standard parametrization,
\bea
\mathcal{U}^{3+2}_{\text{\tiny{PMNS}}}=\prod^{3}_{j>i,i=1}~\mathbb{R}_{ij}\; .
\eea
The multiplication of matrices is to be perfomed from right to left. The rotation matrices may be real or complex. We choose a CP phase parametrization that is consistent with that employed in~\cite{Kopp:2011qd}. Under this convention the matrices $\mathbb{R}_{12},\,\mathbb{R}_{13},\,\mathbb{R}_{15},\,\mathbb{R}_{34}$ and $\mathbb{R}_{35}$ carry CP phases. 
\par
The sterile neutrino matrix elements in this convention are
\bea
U_{e 4}&=&\cos\theta_{15} \sin\theta_{14}\;,\\
U_{e 5}&=&e^{-i\delta_{!5}}\sin\theta_{15}\;,\\
U_{\mu 4}&=&\cos\theta_{14} \cos\theta_{25} \sin\theta_{24}-e^{i\delta_{15}} \sin\theta_{14} \sin\theta_{15} \sin\theta_{25}\;,\\
U_{\mu 5}&=&\cos\theta_{15} \sin\theta_{25}\;.
\eea
\par
The sterile-neutrino angles ($\theta_{14},\,\theta_{15},\,\theta_{24},\,\theta_{25}$) and phase $\delta_{15}$ can in principle be extracted from SBL measurements and specifically the global fits of Table \ref{globalfit}. 
\par
The active-neutrino matrix elements pertinent to our study come out to be
\bea
U_{e 2}&=&\cos\theta_{13} \cos\theta_{14} \cos\theta_{15} e^{-i \delta_{12}} \sin\theta_{12}\;,\\
U_{e 3}&=&\cos\theta_{14} \cos\theta_{15} e^{-i \delta_{13}} \sin\theta_{13}\;,\\
U_{\mu 2}&=&\cos\theta_{12} \cos\theta_{23} \cos\theta_{24} \cos\theta_{25} +  e^{-i \delta_{12}}
   \sin\theta_{12} (\cos\theta_{13} (-\cos\theta_{25} \sin\theta_{14} \sin\theta_{24} - \cos\theta_{14} e^{i \delta_{15}} \sin\theta_{15} \sin\theta_{25}) \nn\\
   &-& \cos\theta_{24} \cos\theta_{25} e^{i \delta_{13}} \sin\theta_{13} \sin\theta_{23})\; ,\\
U_{\mu 3}&=&\cos\theta_{13} \cos\theta_{24} \cos\theta_{25} \sin\theta_{23}-e^{-i \delta_{13}} \sin\theta_{13} (\cos\theta_{25} \sin\theta_{14} \sin\theta_{24}+\cos\theta_{14} e^{i \delta_{15}} \sin\theta_{15} \sin\theta_{25})\;.
\eea
The effective phases $\delta_1$ and $\delta_2$ can be related in principle to the `fundamental' CP phases $\delta_{12}$ and $\delta_{13}$  using the above relations. 
\end{appendix}


\end{document}